\documentclass[12pt]{article}
\usepackage{amsmath}
\usepackage{graphicx}
\usepackage{enumerate}
\usepackage{natbib}
\usepackage{url} % not crucial - just used below for the URL 
\usepackage{binomexp}
\usepackage{mathtools}
\usepackage{amsfonts}
\usepackage{mathrsfs}
\RequirePackage{subcaption}
\usepackage{bbm}
\usepackage{dsfont}
\usepackage{longtable}
\usepackage{yhmath}
\usepackage{bm}
\usepackage{amssymb}
\usepackage{amsbsy}
\usepackage[nodisplayskipstretch]{setspace}
\usepackage{tikz}
\usepackage{float}
\usepackage{stmaryrd}
\usepackage{paralist}
\usepackage{multirow}
\usepackage{threeparttable}
\usepackage{soul}
\usepackage{comment}
\usepackage{algorithm,algpseudocode}
\makeatletter
\newcommand{\algmargin}{\the\ALG@thistlm}
\makeatother
\newlength{\whilewidth}
\algdef{SE}[parWHILE]{parWhile}{EndparWhile}[1]
{\parbox[t]{\dimexpr\linewidth-\algmargin}{%
		\hangindent\whilewidth\strut\algorithmicwhile\ #1\ \algorithmicdo\strut}}{\algorithmicend\ \algorithmicwhile}%
\algnewcommand{\parState}[1]{\State%
	\parbox[t]{\dimexpr\linewidth-\algmargin}{\strut #1\strut}}

\newcommand{\bfa}{\bm{a}}
\newcommand{\bfb}{\bm{b}}
\newcommand{\bfc}{\bm{c}}

\newcommand{\bfe}{\bm{e}}

\newcommand{\bfo}{\bm{o}}

\newcommand{\bfr}{\bm{r}}

\newcommand{\bfx}{\bm{x}}

\newcommand{\bfbeta}{\bm{\beta}}
\newcommand{\bfgamma}{\bm{\gamma}}

\newcommand{\bfepsilon}{\bm{\epsilon}}

\newcommand{\bftau}{\bm{\tau}}

\newcommand{\bfxi}{\bm{\xi}}

\newcommand{\bfPi}{\bm{\Pi}}

\newcommand{\rmI}{\mathrm{I}}
\newcommand{\rmA}{\mathrm{A}}

\newcommand{\rmT}{\mathrm{T}}

\newcommand{\adj}{\mathrm{adj}}
\newcommand{\ancova}{\mathrm{ancova}}

\newcommand{\diag}{\mathrm{diag}}
\newcommand{\se}{\mathrm{se}}
\newcommand{\Cov}{\mathbf{Cov}}

\newcommand{\CR}{\mathrm{CR}}
\newcommand{\HC}{\mathrm{HC}}
\newcommand{\WATE}{\mathrm{WATE}}
\newcommand{\OWTE}{\mathrm{OWTE}}
\newcommand{\AWATE}{\mathrm{AWATE}}
\newcommand{\OAWTE}{\mathrm{OAWTE}}

\newcommand{\bfA}{\bm{A}}

\newcommand{\bfC}{\bm{C}}
\newcommand{\bfD}{\bm{D}}
\newcommand{\bfE}{\bm{E}}
\newcommand{\bfG}{\bm{G}}

\newcommand{\bfI}{\bm{I}}

\newcommand{\bfQ}{\bm{Q}}

\newcommand{\bfV}{\bm{V}}

\newcommand{\bfX}{\bm{X}}

\newcommand{\bbP}{\mathbb{P}}
\newcommand{\bbR}{\mathbb{R}}

\newcommand{\bbone}{\mathbbm{1}}

\newcommand{\calA}{\mathcal{A}}

\newcommand{\calC}{\mathcal{C}}

\newcommand{\calI}{\mathcal{I}}

\newcommand{\calN}{\mathcal{N}}

\newcommand{\calU}{\mathcal{U}}

\newcommand{\calX}{\mathcal{X}}
\newcommand{\calY}{\mathcal{Y}}

\newcommand{\sumi}{\sum_{i=1}^I}
\newcommand{\sumj}{\sum_{j=1}^J}

\newcommand{\sumk}{\sum_{k=1}^{N_{ij}}}
\newcommand{\suma}{\sum_{a\in\calA}}

\usepackage[colorlinks=true,linkcolor=red,citecolor=blue]{hyperref}

\newcommand\bovermat[2]{%
	\makebox[0pt][l]{$\smash{\overbrace{\phantom{%
					\begin{matrix}#2\end{matrix}}}^{\text{#1}}}$}#2}

\newtheorem{theorem}{Theorem}
\newtheorem{assumption}{Assumption}
\newtheorem{proposition}{Proposition}
\newtheorem{corollary}{Corollary}

\newtheorem{remark}{Remark}

\newcommand{\blind}{1}

\allowdisplaybreaks

\usetikzlibrary{shapes,decorations,arrows,calc,arrows.meta,fit,positioning}
\tikzset{
	-Latex,auto,node distance =1 cm and 1 cm,semithick,
	state/.style ={ellipse, draw, minimum width = 0.7 cm},
	point/.style = {circle, draw, inner sep=0.04cm,fill,node contents={}},
	bidirected/.style={Latex-Latex,dashed},
	el/.style = {inner sep=2pt, align=left, sloped}
}

\begin{document}

	\def\spacingset#1{\renewcommand{\baselinestretch}%
		{#1}\small\normalsize} \spacingset{1}

	%%%%%%%%%%%%%%%%%%%%%%%%%%%%%%%%%%%%%%%%%%%%%%%%%%%%%%%%%%%%%%%%%%%%%%%%%%%%%%
	
	\if1\blind
	{
		\title{\Large \bf Model-assisted inference for dynamic causal effects in staggered rollout cluster randomized experiments}
		\author{Xinyuan Chen$^{1,\ast}$ and Fan Li$^{2,\dagger}$\vspace{0.2cm}\\
			$^1$Department of Mathematics and Statistics,\\ Mississippi State University, MS, USA\\
			$^2$Department of Biostatistics, Yale School of Public Health, CT, USA\\
			%$^3$Center for Methods in Implementation and Prevention Science,\\ Yale School of Public Health, CT, USA\\
			${}^\ast$xchen@math.msstate.edu ~ ${}^\dagger$fan.f.li@yale.edu}
		\maketitle
	} \fi
	
	\if0\blind
	{
		\bigskip
		\bigskip
		\bigskip
		\begin{center}
			{\Large \bf Model-assisted inference for dynamic causal effects in staggered rollout cluster randomized experiments
			}
		\end{center}
		\medskip
	} \fi
	
	\bigskip
	\begin{abstract}
		Staggered rollout cluster randomized experiments (SR-CREs) involve sequential treatment adoption across clusters, requiring analysis methods that address a general class of dynamic causal effects, anticipation, and non-ignorable cluster-period sizes. Without imposing any outcome modeling assumptions, we study regression estimators using individual data, cluster-period averages, and scaled cluster-period totals, with and without covariate adjustment from a design-based perspective. We establish consistency and asymptotic normality of each estimator under a randomization-based framework and prove that the associated variance estimators are asymptotically conservative in the L\"{o}wner ordering. Furthermore, we conduct a unified efficiency comparison of the estimators and provide recommendations. We highlight the efficiency advantage of using estimators based on scaled cluster-period totals with covariate adjustment over their counterparts using individual-level data and cluster-period averages. Our results rigorously justify linear regression estimators as model-assisted methods to address an entire class of dynamic causal effects in SR-CREs.
	\end{abstract}
	
	\noindent%

	\spacingset{1.75} % DON'T change the spacing!
	
	% \clearpage 
	\section{Introduction}
	
	% P1. Intro and examples
	We study a class of randomized experiments in which clusters are the unit of randomization, with the treatment rolled out over multiple periods. At baseline, each cluster is randomized to a specific treatment adoption time, and clusters that have already begun treatment will continue to receive treatment until the end of the study. We refer to this design as the staggered rollout cluster-randomized experiment (SR-CRE), which is attractive in settings where simultaneous randomization in a single period is not logistically feasible. SR-CREs have also been referred to as cluster-randomized difference-in-differences (DID) in the social sciences \citep{Athey2022} and stepped-wedge cluster-randomized trials in medicine \citep{Hussey2007}, and have become increasingly common. 
	
	% P2. Literature review
	Previous staggered rollout design literature has primarily focused on single-level data, e.g., observational DID for panel data, where the treatment assignment and outcome measurements occur at the same level \citep{Callaway2021,Sun2021}, and staggered rollout individual randomized experiments \citep[SR-IREs,][]{Athey2022,Bojinov2021,Roth2023}. Differentiating the level of assignment from the level of outcome measurements, there is an emerging literature on design-based inference for CREs without staggered rollout. For example, \citet{Schochet2021} proposed design-based estimators for blocked CREs. \citet{Su2021} developed a unified theory for design-based model-assisted inference in CREs under parallel-arm assignments. Differentiating the level of treatment assignment from the level of outcome measurements under staggered rollout, \citet{Schochet2022} studied the statistical power of DID and comparative interrupted time series estimators for panel data from observational studies. \citet{Chen2023} proposed a class of weighted average treatment effect estimands and studied the design-based properties of analysis of covariance (ANCOVA) estimators for stepped wedge CREs. However, these developments assumed away anticipation \citep[][a psychological phenomenon where clusters bring the expected future into the present]{Malani2015} and dynamic causal effects \citep[][treatment effects evolve over length of exposure]{Sun2021}, which cannot be easily ruled out \emph{a priori} in practice. 
	
	% P3. Our contributions
	Focusing on the central question of treatment effect evaluation, this article addresses regression estimators for SR-CREs from a design-based perspective, where the potential outcomes and covariates are assumed fixed, and the randomness arises solely from the randomized treatment adoption times. We propose a class of weighted average treatment effect estimands that account for the unique design features of an SR-CRE. Specifically, we largely relax the no-anticipation assumption \citep{Athey2022,Chen2023,Roth2023} and allow for dynamic causal effects \citep{Sun2021,wang2024achieve}, such that the treatment adoption time is an intrinsic component of the estimands. Similar to \citet{Su2021} and \citet{Chen2023}, we also explicitly address the cluster-period size as an integral part of the estimands and generalize the dynamic DID estimands \citep{Sun2021,Callaway2021} to address differential levels of assignment and data measurement as well as non-ignorable cluster-period sizes. Based on the proposed estimands, we can then define the dynamic causal effects, which compare the summary of potential outcomes between any pair of treatment adoption times in specific calendar periods to address different scientific questions. To our knowledge, this is the first effort to simultaneously address anticipation, dynamic causal effects, and non-ignorable cluster-period sizes in estimations and definitions for SR-CREs. 
	
	% P4. Estimators and asymptotics
	To target the class of estimands, we propose regression estimators using individual-level data, cluster-period averages, and scaled cluster-period totals, with or without adjusting for covariates. Each proposed estimator is model-assisted as it is consistent even if the working regression model is misspecified. We establish the asymptotic properties of each regression estimator under the asymptotic regime in \citet{Middleton2015} and \citet{Chen2023}, where, with a constant number of periods, the number of clusters approaches infinity while the cluster-period sizes are fixed. Each regression estimator is shown to be consistent and asymptotically normal, provided that the cluster weights in each period are reasonably balanced and there are no clusters with dominating weights. We estimate the uncertainty of each regression estimator based on individual-level data via the Liang-Zeger cluster-robust (CR) variance estimator \citep{Liang1986}, and the uncertainty of those based on cluster-period averages and scaled cluster-period totals via the Huber-White heteroskedasticity-consistent (HC) variance estimator \citep{Huber1967,White1980}. We prove that both variance estimators are conservative. The conservativeness of the HC variance estimators also offers a rigorous justification for a similar statement in \citet{Roth2023}, because regressions based on cluster-period averages and scaled cluster-period totals implicitly treat the cluster-period-level outcomes as though they were derived from an SR-IRE. Finally, we provide a unified efficiency comparison across all regression estimators under SR-CREs and make recommendations.
	
	% P5. Summary
	The rest of this article is structured as follows. Section \ref{sec:design} introduces the SR-CREs, including assumptions and causal estimands. Section \ref{sec:est} proposes regression estimators and establishes their design-based properties, as well as the conservativeness of the variance estimators. Section \ref{sec:comp} compares the estimators theoretically. Section \ref{sec:sim} and \ref{sec:data} report results from simulation studies and an illustrative data example. Section \ref{sec:conc} concludes.

	\section{Staggered rollout cluster randomized experiments} \label{sec:design}
	
	Consider an SR-CRE with $I$ clusters (indexed by $i$) and $J$ rollout periods (indexed by $j$). Each cluster has $J+1$ possible treatment adoption times denoted by $\calA=\{1,\ldots,J,\infty\}$. The treatment can be considered absorbing, as a cluster will remain in the treated condition from the start of the treatment until the experiment terminates. If a cluster is randomized to the treatment adoption time $\infty$, it is said to have never received treatment. The experiment becomes one with the stepped wedge design \citep{Hussey2007} if the treatment adoption time $\infty$ is excluded from $\calA$, where all clusters are eventually treated, but it can be more general. However, even in stepped wedge designs, it is useful to conceptualize an imaginary set of clusters that will be randomized to adoption time $\infty$ for ease of defining the component estimands by contrasting with a hypothetical set of control clusters. Thus, we provide a general discussion where $\infty\in\calA$.
	
	The number of clusters to be randomized to treatment adoption time $a$, denoted by $I(a)$ with $I(a)>0$ for all $a\in\calA$ and $\suma I(a)=I$, is pre-specified, but the corresponding cluster indices of these $I(a)$ clusters, denoted by $\calI(a)$ with $|\calI(a)|=I(a)$, will be determined through randomization. Let $A_i=a\in\calA$ denote the treatment adoption time indicator variable for cluster $i$. The SR-CRE can be alternatively viewed as a $(J+1)$-armed cluster randomized experiment. In addition, an SR-CRE can be cross-sectional, where a different group of individuals is included in each cluster-period, or closed-cohort, where the group of individuals in each cluster is determined at baseline and remains unchanged throughout, or open-cohort, which is a mixture of the two. For a given cluster $i$, its cluster-period size $N_{ij}$ can vary with $j$ in a cross-sectional SR-CRE. Figure \ref{fig:SRE} provides a schematic illustration of an SR-CRE.
	
	\vspace*{0.1in}
	\begin{figure}[htbp]
		\setlength{\unitlength}{0.14in} % selecting unit length
		\centering % used for centering Figure
		\begin{picture}(31.5,11) % picture environment with the size (dimensions)
			% 32 length units wide, and 15 units high.
			\setlength\fboxsep{0pt}
			\put(-1,8){\colorbox{blue!25}{\framebox(7.5,1.5){\scriptsize$\{N_{i1},i\in\calI(1)\}$}}}
			\put(7,8){\colorbox{blue!25}{\framebox(7.5,1.5){\scriptsize$\{N_{i2},i\in\calI(1)\}$}}}
			\put(15,8){\colorbox{blue!25}{\framebox(7.5,1.5){\scriptsize$\{N_{i3},i\in\calI(1)\}$}}}
			\put(23,8){\colorbox{blue!25}{\framebox(7.5,1.5){\scriptsize$\{N_{i4},i\in\calI(1)\}$}}}
			\put(31,8){\colorbox{blue!25}{\framebox(7.5,1.5){\scriptsize$\{N_{i5},i\in\calI(1)\}$}}}
			
			\put(-1,6){\framebox(7.5,1.5){\scriptsize$\{N_{i1},i\in\calI(2)\}$}}
			\put(7,6){\colorbox{blue!25}{\framebox(7.5,1.5){\scriptsize$\{N_{i2},i\in\calI(2)\}$}}}
			\put(15,6){\colorbox{blue!25}{\framebox(7.5,1.5){\scriptsize$\{N_{i3},i\in\calI(2)\}$}}}
			\put(23,6){\colorbox{blue!25}{\framebox(7.5,1.5){\scriptsize$\{N_{i4},i\in\calI(2)\}$}}}
			\put(31,6){\colorbox{blue!25}{\framebox(7.5,1.5){\scriptsize$\{N_{i5},i\in\calI(2)\}$}}}
			
			\put(-1,4){\framebox(7.5,1.5){\scriptsize$\{N_{i1},i\in\calI(3)\}$}}
			\put(7,4){\framebox(7.5,1.5){\scriptsize$\{N_{i2},i\in\calI(3)\}$}}
			\put(15,4){\colorbox{blue!25}{\framebox(7.5,1.5){\scriptsize$\{N_{i3},i\in\calI(3)\}$}}}
			\put(23,4){\colorbox{blue!25}{\framebox(7.5,1.5){\scriptsize$\{N_{i4},i\in\calI(3)\}$}}}
			\put(31,4){\colorbox{blue!25}{\framebox(7.5,1.5){\scriptsize$\{N_{i5},i\in\calI(3)\}$}}}
			
			\put(-1,2){\framebox(7.5,1.5){\scriptsize$\{N_{i1},i\in\calI(4)\}$}}
			\put(7,2){\framebox(7.5,1.5){\scriptsize$\{N_{i2},i\in\calI(4)\}$}}
			\put(15,2){\framebox(7.5,1.5){\scriptsize$\{N_{i3},i\in\calI(4)\}$}}
			\put(23,2){\colorbox{blue!25}{\framebox(7.5,1.5){\scriptsize$\{N_{i4},i\in\calI(4)\}$}}}
			\put(31,2){\colorbox{blue!25}{\framebox(7.5,1.5){\scriptsize$\{N_{i5},i\in\calI(4)\}$}}}
			
			\put(-1,0){\framebox(7.5,1.5){\scriptsize$\{N_{i1},i\in\calI(5)\}$}}
			\put(7,0){\framebox(7.5,1.5){\scriptsize$\{N_{i2},i\in\calI(5)\}$}}
			\put(15,0){\framebox(7.5,1.5){\scriptsize$\{N_{i3},i\in\calI(5)\}$}}
			\put(23,0){\framebox(7.5,1.5){\scriptsize$\{N_{i4},i\in\calI(5)\}$}}
			\put(31,0){\colorbox{blue!25}{\framebox(7.5,1.5){\scriptsize$\{N_{i5},i\in\calI(5)\}$}}}
			
			\put(-1,-2){\framebox(7.5,1.5){\scriptsize$\{N_{i1},i\in\calI(\infty)\}$}}
			\put(7,-2){\framebox(7.5,1.5){\scriptsize$\{N_{i2},i\in\calI(\infty)\}$}}
			\put(15,-2){\framebox(7.5,1.5){\scriptsize$\{N_{i3},i\in\calI(\infty)\}$}}
			\put(23,-2){\framebox(7.5,1.5){\scriptsize$\{N_{i4},i\in\calI(\infty)\}$}}
			\put(31,-2){\framebox(7.5,1.5){\scriptsize$\{N_{i5},i\in\calI(\infty)\}$}}
			
			% \put(-1,-2.5){$\underbrace{\hspace{6.5em}}_{pre-rollout}$}
			% \put(7,-2.5){$\underbrace{\hspace{20.25em}}_{rollout}$}
			% \put(31,-2.5){$\underbrace{\hspace{6.5em}}_{post-rollout}$}
			
			\put(-5,-1.5){\scriptsize$\calI(\infty)$}
			\put(-5,0.5){\scriptsize$\calI(5)$}
			\put(-5,2.5){\scriptsize$\calI(4)$}
			\put(-5,4.5){\scriptsize$\calI(3)$}
			\put(-5,6.5){\scriptsize$\calI(2)$}
			\put(-5,8.5){\scriptsize$\calI(1)$}
			
			\put(-6.25,10.5){Clusters}
			\put(0.5,10.5){Period $1$}
			\put(8.5,10.5){Period $2$}
			\put(16.5,10.5){Period $3$}
			\put(24.5,10.5){Period $4$}
			\put(32.5,10.5){Period $5$}
		\end{picture}
		\vspace*{0.25in}
		\caption{An example schematic of an SR-CRE with $I=\suma I(a)$ clusters and $J=5$ rollout periods. There are six possible treatment adoption times, $\calA=\{1,2,3,4,5,\infty\}$. Each row represents a subset of all clusters with a unique treatment adoption time. Each period is of equal length, and we include in each cell all cluster-period sizes ($N_{ij}$) for clusters included in that row during that period. A white cell indicates the control condition, and a shaded cell indicates the treatment condition. 
		}
		% \vspace*{-0.1in}
		\label{fig:SRE}
	\end{figure}
	
	\subsection{Identification assumptions}
	
	We state minimally sufficient identification assumptions below.
	\begin{assumption}[Cluster-level SUTVA] \label{asp:sutva}
		Let $Y_{ijk}(A_i,\bfA_{-i})$ denote the potential outcome for individual $k=1,\ldots,N_{ij}$ from cluster $i$ in period $j$ given all adoption times, where $\bfA_{-i}$ denote the vector of adoption times for all other clusters, then (i) $Y_{ijk}(A_i,\bfA_{-i}) = Y_{ijk}(A_i,\bfA_{-i}^*)$, for all $\bfA_{-i}\neq \bfA_{-i}^*$; and (ii) $Y_{ijk}(A_i,\bfA_{-i}) = Y_{ijk}(A_i^*,\bfA_{-i}^*)$ if $A_i=A_i^*$.
	\end{assumption}
	
	As a common condition, Assumption \ref{asp:sutva} states that there is a single version of the treatment, and individual-level potential outcomes from cluster $i$ in period $j$ only depend on the assignment of cluster $i$. It rules out between-cluster interference and simplifies the potential outcome as $Y_{ijk}(A_i)$. Under Assumption \ref{asp:sutva}, the observed individual outcome is $Y_{ijk}=\sum_{a\in\calA}G_i(a)Y_{ijk}(a)$, where $G_i(a)=\bbone(A_i=a)$. We use $\bfX_{ijk}$ and $\bfC_{ij}$ to denote the $p_x$- and $p_c$-dimensional row vectors of individual and cluster-level covariates in period $j$, respectively. The cluster-level covariates $\bfC_{ij}$ possibly include components of cluster-level summaries. We assume both $\bfX_{ijk}$ and $\bfC_{ij}$ are exogenous; in the case of a cohort study, one can remove the subscript $j$ such that $\bfX_{ik}$ and $\bfC_i$ only involve time-invariant information collected before individual recruitment. 
	
	\begin{assumption}[Randomized staggered rollout]\label{asp:rand}
		Write $\calY$, $\calX$, and $\calC$ as the collections of all potential outcomes and covariates across individuals, clusters, and periods, then 
		\begin{equation*}
			\bbP(\bfA=\bfa|\calY,\calX,\calC)={I\choose I(1),\ldots,I(J),I(\infty)}^{-1},
		\end{equation*}
		where $\bfA=(A_1,\ldots,A_I)^\top$ and $\bfa=(a_1,\ldots,a_I)^\top$ with $\sumi\bbone(A_i=a)=I(a)$.
	\end{assumption}
	
	This assumption states that the treatment adoption times are random. It is a common assumption in the staggered rollout design literature, e.g., \citet[Assumption 1]{Athey2022}, \citet[Assumption 1]{Roth2023}, and \citet[Assumption 3]{Chen2023}, and is typically held by design.
	
	\subsection{Dynamic causal effect estimands}
	
	We pursue the finite-population framework. Under this framework, values of the potential outcomes and covariates are assumed to be fixed, and the source of randomness is solely from the randomization of the treatment adoption time assignments. Under Assumption \ref{asp:sutva}, we define the dynamic weighted average treatment effect (DWATE) estimand, 
	\begin{equation} \label{eq:d-wate}
		\tau_j(a,a')=\overline Y_{\cdot j\cdot}(a)-\overline Y_{\cdot j\cdot}(a'),
	\end{equation}
	where $\overline Y_{\cdot j\cdot}(a)$ is the weighted average potential outcomes across individuals in period $j$, had the adoption time been set to $a$. The DWATE is a contrast in weighted average potential outcomes between different treatment adoption times in period $j=1,\ldots,J$.
	
	Let $w_{ijk}$ denote the non-negative pre-specified individual weight for individual $k$ from cluster $i$ in period $j$, which satisfies the constraint that $\sumi\sumk w_{ijk}>0$ for all $j$, ensuring \eqref{eq:d-wate} is properly defined. We define $w_{ij\cdot}=\sumk w_{ijk}$ as the cluster weight for cluster $i$ in period $j$, $w_{\cdot j\cdot}=\sumi w_{ij\cdot}$ as the total weight for period $j$, $\pi_{ijk}=w_{ijk}/w_{\cdot j\cdot}$ as the normalized individual weight, and $\pi_{ij\cdot}=\sumk\pi_{ijk}=w_{ij\cdot}/w_{\cdot j\cdot}$ as the normalized cluster weight. The DWATE in \eqref{eq:d-wate} can thus be re-expressed using individual-level data as
	\begin{equation} \label{eq:d-wate-I}
		\tau_j(a,a')=\sumi\sumk \pi_{ijk}\{Y_{ijk}(a)-Y_{ijk}(a')\},
	\end{equation}
	with $\overline Y_{\cdot j\cdot}(a)=\sumi\sumk \pi_{ijk}Y_{ijk}(a)$; and using the cluster-period averages as
	\begin{equation} \label{eq:d-wate-A}
		\tau_j(a,a')=\sumi \pi_{ij\cdot}\{\overline Y_{ij\cdot}(a)-\overline Y_{ij\cdot}(a')\},
	\end{equation}
	where $\overline Y_{ij\cdot}(a)=\sumk w_{ijk}Y_{ijk}(a)/w_{ij\cdot}$ for $a\in\calA$; and further using the scaled cluster-period totals,
	\begin{equation} \label{eq:d-wate-T}
		\tau_j(a,a')=I^{-1}\sumi\{\widetilde Y_{ij\cdot}(a)-\widetilde Y_{ij\cdot}(a')\},
	\end{equation}
	where $\widetilde Y_{ij\cdot}(a)=I\pi_{ij\cdot}\overline Y_{ij\cdot}(a)$ for $a\in\calA$. Different estimators for $\tau_j(a,a')$ using individual-level data, cluster-period averages, and scaled cluster-period totals can then be motivated by the estimand re-expressions in \eqref{eq:d-wate-I} - \eqref{eq:d-wate-T}, and will be addressed later. Finally, although one can specify arbitrary individual weights $w_{ijk}$, the two most interpretable choices have been discussed in prior work \citep{kahan2022estimands,Chen2023}. These include (i) setting $w_{ijk}=1$ ($\pi_{ijk}=(\sumj N_{ij})^{-1}$) to assign equal weight to each individual such that $\tau_j(a,a')$ is an individual average, and (ii) setting $w_{ijk}=N_{ij}^{-1}$ ($\pi_{ijk}=(IN_{ij})^{-1}$) to assign equal weight to each cluster during each period such that $\tau_j(a,a')$ becomes a cluster average.

	\begin{table}[htbp] %***
		\caption{Causal interpretations of the DWATE estimand $\tau_j(a,a')$ with $j=1,\ldots,J$ and assuming $a,a'\in\calA$ and $a<a'$.}\label{tab:d-wate-interpretations}
		\vspace{-0.15in}
		\begin{center}
			\resizebox{\linewidth}{!}{
				\begin{tabular}{cl} 
					\hline %***5truept
					\textbf{Relationships} & \multicolumn{1}{c}{\textbf{Causal interpretations}} \\
					\hline
					$j<a<a'$ & The anticipation effect of delaying the treatment adoption time \\
					& from $a$ to $a'$, evaluated in calendar period $j$.\\
					$a\leq j<a'$ & The contrast effect of adopting treatment for $j-a+1$ periods compared \\
					& to anticipating to adopt at time $a'$, evaluated in calendar period $j$.\\
					$a<a'\leq j$ & The duration effect of lag $(a'-a)$ that compares treatment adoption\\
					& times $a$ and $a'$, evaluated in calendar period $j$.\\
					\hline
				\end{tabular}
			}
		\end{center}
	\end{table}
	\vspace{-0.2in}
	
	The definition of $\tau_j(a,a')$ in \eqref{eq:d-wate} implies that $\tau_j(a',a)=\overline Y_{\cdot j\cdot}(a')-\overline Y_{\cdot j\cdot}(a)=-\tau_j(a,a')$ and $\tau_j(a,a)=0$. Therefore, it is sufficient to focus on $\tau_j(a,a')$'s with $a<a'$. Depending on the comparative relationship between period $j$ and treatment adoption times $a$ and $a'$, the DWATE estimand $\tau_j(a,a')$ has the following causal interpretations summarized in Table \ref{tab:d-wate-interpretations}. Causal interpretations of $\tau_j(a,a')$ are straightforward for $j<a<a'$ and $a<a'\leq j$, as they capture fine-resolution anticipation and duration effects, respectively. Of note, there are no anticipation effects when $j=J$ or $a=1$, and naturally, the anticipation effects are assumed to exist only for eventually treated clusters. For $a\leq j<a'$, causal interpretations of $\tau_j(a,a')$ could be more subtle for a general $a'$ because of the anticipation component brought by the relationship $j<a'$. Nevertheless, we provide a discussion on the general $\tau_j(a,a')$ for theoretical interests and completeness of the subsequent development. To focus ideas, the causal interpretations in this scenario can be more streamlined by setting $a'=\infty$, allowing us to draw comparisons to the average potential outcomes had the clusters never been treated (an imaginary counterfactual control condition). In this case, we can define $\WATE_j(a)=\tau_j(a,\infty)$ for $a\leq j$ as the weighted average treatment effect (WATE) of treatment adoption time $a$ measured for the population in calendar period $j$. The WATE is analogous to but differs from the average treatment effect on the treated under non-random treatment adoption times \citep{Callaway2021,Sun2021}, as the latter is defined for the treated group. Furthermore, when setting $a'=\infty$, we can also define $\AWATE_j(a)=\tau_j(a,\infty)$ for $j<a$ as the anticipated WATE (AWATE) of treatment adoption time $a$ measured for the population in calendar period $j$ \citep{Abbring2003,Malani2015}. Additional special cases of our estimands can be found in Remark S1 in Section S1 of the Online Supplement.
	
	The DWATE $\tau_j(a,a')$ can be used in constructing different collections of practically useful summary estimands in the form of
	\begin{align} \label{eq:gen-estimand}
		\theta=\sumj\sum_{a<a'}b_{\theta,j}(a,a')\tau_j(a,a'),
	\end{align}
	which is a weighted sum of $\tau_j(a,a')$ with $b_{\theta,j}(a,a')\in\bbR$ being user-defined weights. For example, with $\tau_j(a,\infty)=\WATE_j(a)$ ($a\leq j$), we can define a grand average estimand over adoption and calendar times as $\OWTE^{sim} = \sumj\sum_{a:a\leq j}w_{\cdot j\cdot}I(a)\WATE_j(a)/\allowbreak\sumj\sum_{a:a\leq j}w_{\cdot j\cdot}I(a)$. Similarly, Using $\tau_j(a,\infty)=\AWATE_j(a)$ ($j<a$) as building blocks, a grand average estimand with a focus on anticipation effects is given by $\OAWTE^{sim} = \sum_{j=1}^{J-1}\sum_{a:j<a\leq J}\allowbreak w_{\cdot j\cdot}I(a)\AWATE_j(a)/\sum_{j=1}^{J-1}\sum_{a:j<a\leq J}w_{\cdot j\cdot}I(a)$. We provide four classes of summary estimands to address different scientific questions in Section S1 of the Online Supplement.

	\section{Regression-based causal effects estimators} \label{sec:est}
	
	\subsection{Preliminaries} \label{sec:prelim}
	
	We present preliminaries to provide the context in which the asymptotic properties of the estimators for $\tau_j(a,a')$ will be discussed. We adopt the asymptotic regime in \citet{Middleton2015}, \citet{Li2017}, and \citet{Chen2023}, where an increasing sequence of finite populations is considered with the total number of clusters, $I\rightarrow\infty$, while the number of rollout periods, $J$, remains fixed. In addition, we assume that the cluster size $N_{ij}$ remains fixed as $I$ grows. 
	
	Let $\bfxi_{i\cdot}(a)=(\xi_{i1\cdot}(a),\ldots,\xi_{iJ\cdot}(a))^\top$ for $a\in\calA$ and $i=1,\ldots,I$ denote general cluster-level potential outcomes with zero averages over clusters. Define the individual centered potential outcome $\epsilon_{ijk}(a)=Y_{ijk}(a)-\overline Y_{\cdot j\cdot}(a)$, and then examples of $\xi_{ij\cdot}(a)$ include the cluster-period average centered potential outcome, $\overline\epsilon_{ij\cdot}(a)=\overline Y_{ij\cdot}(a)-\overline Y_{\cdot j\cdot}(a)$, and the scaled cluster-period total, $\widetilde\epsilon_{ij\cdot}(a)=\widetilde Y_{ij\cdot}(a)-I\pi_{ij\cdot}\overline Y_{\cdot j\cdot}(a)$. We then define the following matrices to streamline the exposition of the theoretical properties of the estimators, where
	\begin{align*} %\label{eq:variance-c}
		\bfV_c(\bfxi_{i\cdot})(a,a')=\sumi\left\{\frac{\bfxi_{i\cdot}(a)\bfxi_{i\cdot}(a)^\top}{I(a)}+\frac{\bfxi_{i\cdot}(a')\bfxi_{i\cdot}(a')^\top}{I(a')}\right\}
	\end{align*}
	and $\bfV(\bfxi_{i\cdot})(a,a')=\bfV_c(\bfxi_{i\cdot})(a,a')-I^{-1}\sumi\left\{\bfxi_{i\cdot}(a)-\bfxi_{i\cdot}(a')\right\}\left\{\bfxi_{i\cdot}(a)-\bfxi_{i\cdot}(a')\right\}^\top$. For two treatment adoption times $a$ and $a'$, $\bfV_c(\bfxi_{i\cdot})(a,a')$ and $\bfV(\xi_{i\cdot})(a,a')$ are covariance matrices of the difference in means of $\{\bfxi_{i\cdot}(a),\bfxi_{i\cdot}(a')\}_{i=1}^I$ multiplied by $(I-1)$ in completely randomized experiments with and without the constant treatment effect assumption. In a more granular fashion, examining the $(j,j)$ entry of $\bfV_c(\bfxi_{i\cdot})(a,a')$ and $\bfV(\bfxi_{i\cdot})(a,a')$ gives us 
	\begin{align*} %\label{eq:variance-c}
		\bfV_c(\bfxi_{i\cdot})(a,a')_{(j,j)}=\sumi\left\{\frac{\xi_{ij\cdot}(a)^2}{I(a)}+\frac{\xi_{ij\cdot}(a')^2}{I(a')}\right\}
	\end{align*}
	and $\bfV(\bfxi_{i\cdot})(a,a')_{(j,j)}=\bfV_c(\bfxi_{i\cdot})(a,a')_{(j,j)}-I^{-1}\sumi\left\{\xi_{ij\cdot}(a)-\xi_{ij\cdot}(a')\right\}^2$, which are variances of the difference in means of $\{\xi_{ij\cdot}(a),\xi_{ij\cdot}(a')\}_{i=1}^I$ multiplied by $(I-1)$ in completely randomized experiments with and without the constant treatment effect assumption. 
	
	\subsection{Estimators using individual-level data} \label{sec:est-ind}
	
	\subsubsection{Without covariates}
	
	We can estimate $\tau_j(a,a')$ for $j=1,\ldots,J$ and assuming $a,a'\in\calA$ and $a< a'$ using individual-level data, motivated by the expression in \eqref{eq:d-wate-I}. Consider the following working regression model,
	\begin{equation}  \label{eq:model-I}
		Y_{ijk} = \suma\beta_j(a)G_i(a)+e_{ijk},
	\end{equation}
	where $e_{ijk}$ is the individual-level random noise assumed to have a mean-zero distribution and is independent across individuals. We can obtain an estimator for $\tau_j(a,a')$ via the weighted least squares (WLS) fit of \eqref{eq:model-I} with individual weight $\pi_{ijk}$, in the form of $\widehat\tau_{j,\rmI}(a,a')=\widehat\beta_{j,\rmI}(a)-\widehat\beta_{j,\rmI}(a')$. Here, $\widehat\beta_{j,\rmI}(a)=\overline y_{\cdot j\cdot}(a)=\sumi G_i(a) \pi_{ij\cdot}\overline Y_{ij\cdot}/\pi_{\cdot j\cdot}(a)$ and $\pi_{\cdot j\cdot}(a)=\sumi G_i(a)\pi_{ij\cdot}$. The estimator $\widehat\tau_{j,\rmI}(a,a')$ is equivalent to the difference between weighted averages of observed outcomes in clusters randomized to treatment adoption times $a$ and $a'$, and thus can also be labeled as the difference-in-means estimator. When $|\calA|=2$, this estimator reduces to the regression estimators without adjusting for covariates under parallel-arm CREs \citep{Schochet2021,Su2021}. We use the subscript `$\rmI$' to clarify that the estimator is obtained using individual-level data. 
	
	Let $\Cov\{\widehat\bftau_{\rmI}(a,a')\}$ denote the covariance matrix of $\widehat\bftau_{\rmI}(a,a')=(\widehat\tau_{1,\rmI}(a,a'),\ldots,\widehat\tau_{J,\rmI}(a,a'))^\top$ and $\se^2\{\widehat\tau_{j,\rmI}(a,a')\}$ denote the marginal variance of $\widehat\tau_{j,\rmI}(a,a')$, where $\se^2\{\widehat\tau_{j,\rmI}(a,a')\}$ is the $(j,j)$ entry of $\Cov\{\widehat\bftau_{\rmI}(a,a')\}$. Both $\Cov\{\widehat\bftau_{\rmI}(a,a')\}$ and $\se^2\{\widehat\tau_{j,\rmI}(a,a')\}$ can be estimated by the CR variance estimator \citep{Liang1986}, denoted by $\widehat\Cov_\CR\{\widehat\bftau_{\rmI}(a,a')\}$ and $\widehat\se_\CR^2\{\widehat\tau_{j,\rmI}(a,a')\}$, respectively. For a WLS fit of \eqref{eq:model-I}, let $\bfD_{i,\rmI}$ and $\bfPi_{i,\rmI}$ respectively denote the design matrix and diagonal weight matrix for cluster $i$. We first obtain the following CR variance estimator for the vector of the regression coefficients, 
	\begin{align*} %\label{eq:CRSE-beta-I}
		\bfE_{\rmI}=\left(\sum_{i=1}^I\bfD_{i,\rmI}^\top\bfPi_{i,\rmI}\bfD_{i,\rmI}\right)^{-1}\left(\sumi\bfD_{i,\rmI}^\top\bfPi_{i,\rmI}\widehat{\bfe}_{i,\rmI}\widehat{\bfe}_{i,\rmI}^\top\bfPi_{i,\rmI}\bfD_{i,\rmI}\right)\left(\sum_{i=1}^I\bfD_{i,\rmI}^\top\bfPi_{i,\rmI}\bfD_{i,\rmI}\right)^{-1},
	\end{align*}
	where $\widehat\bfe_{i,\rmI}$ is the $N_{i\cdot}$-dimensional vector of residuals from the WLS fit of \eqref{eq:model-I}, with $N_{i\cdot}=\sumj N_{ij}$, $\bfD_{i,\rmI}$ is of dimensions $N_{i\cdot}\times J(J+1)$ with $\bfD_{i,\rmI}=(\bfD_{i1,\rmI}^\top\cdots \bfD_{iJ,\rmI}^\top)^\top$ and the $k$-th row of $\bfD_{ij,\rmI}$ being $\bfG_i(\calA)^\top\otimes\pmb\bbone_{ijk}^\top$, and the $j$-th diagonal block of $\bfPi_{i,\rmI}$ is $\bfPi_{ij,\rmI}=\diag(\pi_{ij1},\ldots,\pi_{ijN_{ij}})$. Here, $\bfG_i(\calA)=(G_i(1),\ldots,G_i(J),G_i(\infty))^\top$, $\pmb\bbone_{ijk}=(\bbone_{ijk,1}\ldots,\bbone_{ijk,J})^\top$ with $\bbone_{ijk,j'}=1$ if $j=j'$ and $\bbone_{ijk,j'}=0$ otherwise for $j'=1,\ldots,J$, and `$\otimes$' denotes the Kronecker product such that $\bfG_i(\calA)^\top\otimes\pmb\bbone_{ijk}^\top=(G_i(1)\bbone_{ijk}^\top,\ldots,G_i(J)\bbone_{ijk}^\top,\allowbreak G_i(\infty)\bbone_{ijk}^\top)$. We can write $\widehat\bftau_{\rmI}(a,a')=\bfQ(a,a')\widehat\bfbeta_{\rmI}$, where $\widehat\bfbeta_{\rmI}=(\widehat\bfbeta_{\rmI}(1)^\top,\ldots,\widehat\bfbeta_{\rmI}(J)^\top,\widehat\bfbeta_{\rmI}(\infty)^\top)^\top$, with $\widehat\bfbeta_{\rmI}(a)=(\widehat\beta_{1,\rmI}(a),\ldots,\widehat\beta_{J,\rmI}(a))^\top$ for $a\in\calA$, and $\bfQ(a,a')$ is a $J\times J(J+1)$-dimensional known coefficient matrix, e.g.,
	\begin{align*}
		\bfQ(1,2)=\begin{pmatrix}
			\bfI_{J\times J} & -\bfI_{J\times J} & \bovermat{$J-1$}{\bm{0}_{J\times J} & \cdots & \bm{0}_{J\times J}}
		\end{pmatrix}
	\end{align*}
	and
	\begin{align*}
		\bfQ(2,3)=\begin{pmatrix}
			\bm{0}_{J\times J} & \bfI_{J\times J} & -\bfI_{J\times J} & \bovermat{$J-2$}{\bm{0}_{J\times J} & \cdots & \bm{0}_{J\times J}}
		\end{pmatrix}.
	\end{align*}
	Then, the CR variance estimator for $\Cov\{\widehat\bftau_{\rmI}(a,a')\}$ is $\widehat\Cov_\CR\{\widehat\bftau_{\rmI}(a,a')\}=\bfQ(a,a')\bfE_\rmI\bfQ(a,a')^\top$, and the CR variance estimator for $\se^2\{\widehat\tau_{j,\rmI}(a,a')\}$ is
	\begin{align*}  %\label{eq:CRSE-I}
		\widehat \se_\CR^2\{\widehat\tau_{j,\rmI}(a,a')\}
		&=\widehat\Cov_\CR\{\widehat\bftau_{\rmI}(a,a')\}_{(j,j)}\displaybreak[0]\\
		&=\sumi\frac{G_i(a)}{\pi_{\cdot j\cdot}(a)^2}\left(\sumk \pi_{ijk}\widehat e_{ijk,\rmI}\right)^2+\sumi\frac{G_i(a')}{\pi_{\cdot j\cdot}(a')^2}\left(\sumk \pi_{ijk}\widehat e_{ijk,\rmI}\right)^2.
	\end{align*}
	The following result summarizes the theoretical properties of $\widehat\bftau_{\rmI}(a,a')$ and $\widehat \Cov_\CR\{\widehat\bftau_{\rmI}(a,a')\}$.
	
	\begin{theorem} \label{thm:consistency-AN-I}
		Under Assumptions \ref{asp:sutva} and \ref{asp:rand}, and regularity conditions (C1) and (C2) in Section S2 of the Online Supplement, for $a<a'\in\calA$, $\widehat\bftau_{j,\rmI}(a,a')=\bftau(a,a')+\bfo_\bbP(1)$; if we further assume $\max_i\pi_{ij\cdot}=o(I^{-2/3})$ for $j=1,\ldots,J$ and $\bfV(\widetilde\bfepsilon_{i\cdot})(a,a')\nrightarrow \bm{0}_{J\times J}$, then $I^{1/2}\{\widehat\bftau_{\rmI}(a,a')-\bftau(a,a')\}\xrightarrow{d}\calN(\bm{0}_J,\bfV(\widetilde\bfepsilon_{i\cdot})(a,a'))$, where $\widetilde\bfepsilon_{i\cdot}=(\widetilde\epsilon_{i1\cdot}(a),\ldots,\widetilde\epsilon_{iJ\cdot}(a))^\top$, and $I\times\widehat \Cov_\CR\{\widehat\bftau_{\rmI}(a,a')\}=\bfV_c(\widetilde\bfepsilon_{i\cdot})(a,a')+\bfo_\bbP(1)$.
	\end{theorem}
	
	Here, $\bfV(\widetilde\bfepsilon_{i\cdot})(a,a')\nrightarrow \bm{0}_{J\times J}$ means that $\bfV(\widetilde\bfepsilon_{i\cdot})(a,a')$ converges to a limit not equal $\bm{0}_{J\times J}$. Theorem \ref{thm:consistency-AN-I} establishes the consistency and asymptotic normality of $\widehat\bftau_{\rmI}(a,a')$. The regularity condition on the order of $\pi_{ij\cdot}$ is slightly stronger for establishing the asymptotic normality. Theorem \ref{thm:consistency-AN-I} also shows that the CR variance estimator $\widehat \Cov_\CR\{\widehat\bftau_{\rmI}(a,a')\}$ is conservative, in the L\"{o}wner ordering, for the true covariance matrix because of the difference between $\bfV_c(\widetilde\bfepsilon_{i\cdot})(a,a')$ and $\bfV(\widetilde\bfepsilon_{i\cdot})(a,a')$ is non-negative definite, and is inestimable in general and only equals a zero matrix when $\widetilde\bfepsilon_{i\cdot}(a)-\widetilde\bfepsilon_{i\cdot}(a')=0$ or $\overline Y_{ij\cdot}(a)-\overline Y_{ij\cdot}(a')=\tau_j(a,a')$ for all $i$.
	
	\subsubsection{With covariates}
	
	Covariate adjustment can be incorporated into \eqref{eq:model-I} following \citet{Lin2013} to potentially improve estimation efficiency, leading to the fully-interacted regression model:
	\begin{equation}  \label{eq:model-I-cov}
		Y_{ijk} = \suma\beta_j(a)G_i(a)+\suma G_i(a)\bfX_{ijk}^c\bfgamma_j(a)+e_{ijk},
	\end{equation}
	where the centered individual-level covariate vector $\bfX_{ijk}^c=\bfX_{ijk}-\overline\bfX_{\cdot j\cdot}$ with $\overline{\bfX}_{\cdot j\cdot}=\sumi \pi_{ij\cdot}\overline{\bfX}_{ij\cdot}$ and $\overline{\bfX}_{ij\cdot}=\sumk \pi_{ijk}\bfX_{ijk}\allowbreak/\pi_{ij\cdot}$, and $\bfgamma_j(a)$ is the adoption-time-specific parameter vector. Similar to $\widehat\beta_{j,\rmI}(a)$, with individual weight $\pi_{ijk}$, we can obtain WLS estimates $\widehat\beta_{j,\rmI}^\adj(a)$ and $\widehat\bfgamma_{j,\rmI}(a)$ via fitting \eqref{eq:model-I-cov}. A covariate-adjusted estimator for $\tau_j(a,a')$ can be obtained in the form of $\widehat\tau_{j,\rmI}^\adj(a,a')=\widehat\beta_{j,\rmI}^\adj(a)-\widehat\beta_{j,\rmI}^\adj(a')$, with $\widehat\beta_{j,\rmI}^\adj(a)=\overline y_{\cdot j\cdot}(a)-\overline\bfx_{\cdot j\cdot}^c(a)\widehat\bfgamma_{j,\rmI}(a)$ where $\overline\bfx_{\cdot j\cdot}^c(a)=\sumi \pi_{ij\cdot}G_i(a)\overline\bfX_{ij\cdot}^c/\allowbreak\pi_{\cdot j\cdot}(a)$ and $\overline\bfX_{ij\cdot}^c=\sumk \pi_{ijk}\bfX_{ijk}^c/\pi_{ij\cdot}$. The covariate-adjusted estimator $\widehat\tau_{j,\rmI}^\adj(a,a')$ can be viewed as a multi-arm clustered extension of the regression estimator using a covariate-treatment indicator interaction model, as presented in \citet{Lin2013}. 
	
	To obtain the CR variance estimator $\widehat\Cov_\CR\{\widehat\bftau_{\rmI}^\adj(a,a')\}$, we have
	\begin{align*}
		\bfE_{\rmI}^\adj&=\left(\sum_{i=1}^I(\bfD_{i,\rmI}^\adj)^\top\bfPi_{i,\rmI}\bfD_{i,\rmI}^\adj\right)^{-1}\left(\sumi(\bfD_{i,\rmI}^\adj)^\top\bfPi_{i,\rmI}\widehat{\bfe}_{i,\rmI}^\adj(\widehat{\bfe}_{i,\rmI}^\adj)^\top\bfPi_{i,\rmI}\bfD_{i,\rmI}^\adj\right)\displaybreak[0]\\
		&\qquad\times\left(\sum_{i=1}^I(\bfD_{i,\rmI}^\adj)^\top\bfPi_{i,\rmI}\bfD_{i,\rmI}^\adj\right)^{-1},
	\end{align*}
	where $\widehat\bfe_{i,\rmI}^\adj$ is the vector of residuals from the WLS fit of \eqref{eq:model-I-cov}, $\bfD_{i,\rmI}^\adj$ is of dimensions $N_{i\cdot}\times J(1+p_x)(J+1)$ with $\bfD_{i,\rmI}^\adj=((\bfD_{i1,\rmI}^\adj)^\top\cdots (\bfD_{iJ,\rmI}^\adj)^\top)^\top$ and the $k$-th row of $\bfD_{ij,\rmI}^\adj$ being $(\bfG_i(\calA)^\top\otimes\pmb\bbone_{ijk}^\top,\bfG_i(\calA)^\top\otimes(\pmb\bbone_{ijk}^\top\otimes\bfX_{ijk}^c))$. The CR variance estimator for $\Cov\{\widehat\bftau_{\rmI}^\adj(a,a')\}$ is $\widehat\Cov_\CR\{\widehat\bftau_{\rmI}^\adj(a,a')\}=\bfQ(a,a')\bfE_{\rmI,[1\rightarrow J(J+1),1\rightarrow J(J+1)]}^\adj\bfQ(a,a')^\top$, where $\bfE_{\rmI,[1\rightarrow J(J+1),1\rightarrow J(J+1)]}^\adj$ is the submatrix consisting of the first $J(J+1)$ columns and $J(J+1)$ rows of $\bfE_{\rmI}^\adj$. Let $\bfgamma_{j,\rmI}(a)$ denote the WLS coefficient vectors obtained using individual-level data if the full set of potential outcomes is available. That is, regress $Y_{ijk}(a)$ on $(1,\bfX_{ijk}^c)$ with weight $\pi_{ijk}$ assuming $Y_{ijk}(a)$ is available for each individual $k$ from cluster $i$ in period $j$. We further define the scaled cluster-period totals for non-centered and centered covariates as $\widetilde\bfX_{ij\cdot}=I\pi_{ij\cdot}\overline\bfX_{ij\cdot}$ and $\widetilde\bfX_{ij\cdot}^c=I\pi_{ij\cdot}\overline\bfX_{ij\cdot}^c$, respectively. The theoretical properties of $\widehat\bftau_{\rmI}^\adj(a,a')$ and $\widehat\Cov_\CR\{\widehat\bftau_{\rmI}^\adj(a,a')\}$ are given in Theorem \ref{thm:consistency-AN-I-cov}.
	\begin{theorem} \label{thm:consistency-AN-I-cov}
		Define $r_{ij\cdot}(a)=\widetilde\epsilon_{ij\cdot}(a)-\widetilde\bfX_{ij\cdot}^c\bfgamma_{j,\rmI}(a)$ and $\bfr_{i\cdot}(a)=(r_{i1\cdot}(a),\ldots,r_{iJ\cdot}(a))^\top$. Under Assumptions \ref{asp:sutva} and \ref{asp:rand}, and regularity conditions (C1) - (C3) in Section S2 of the Online Supplement, for $a<a'\in\calA$, $\widehat\bftau_{\rmI}^\adj(a,a')=\bftau(a,a')+\bfo_\bbP(1)$; if further $\max_i \pi_{ij\cdot}=o(I^{-2/3})$ for $j=1,\ldots,J$ and $\bfV(\bfr_{i\cdot})(a,a')\nrightarrow \bm{0}_{J\times J}$, then $I^{1/2}\{\widehat\bftau_{\rmI}^\adj(a,a')-\bftau(a,a')\}\xrightarrow{d}\calN(\bm{0}_J,\bfV(\bfr_{i\cdot})(a,a'))$, and $I\times\widehat \Cov_\CR\{\widehat\bftau_{\rmI}^\adj(a,a')\}=\bfV_c(\bfr_{i\cdot})(a,a')+\bfo_\bbP(1)$.
	\end{theorem}
	
	Theorem \ref{thm:consistency-AN-I-cov} establishes the consistency and asymptotic normality of $\widehat\bftau_{\rmI}^\adj(a,a')$, and the conservativeness of the CR variance estimator $\widehat \Cov_\CR\{\widehat\bftau_{\rmI}^\adj(a,a')\}$, in the L\"{o}wner ordering. For the marginal variance $\se^2\{\widehat\tau_{j,\rmI}^\adj(a,a')\}$, the residual $r_{ij\cdot}(a)$ in Theorem \ref{thm:consistency-AN-I-cov} is of the form $\widetilde\epsilon_{ij\cdot}(a)-\widetilde\bfX_{ij\cdot}^c\bfgamma$, which suggests that the optimal choice of $\bfgamma$ should be the coefficient of $\widetilde\bfX_{ij\cdot}^c$ in the ordinary least squares (OLS) fit of $\widetilde\epsilon_{ij\cdot}(a)$ on $\widetilde\bfX_{ij\cdot}^c$ based on the scaled cluster-period totals, rather than $\bfgamma_{j,\rmI}(a)$ using individual-level data. Therefore, it is possible that $\widehat\tau_{j,\rmI}^\adj(a,a')$ is less efficient than $\widehat\tau_{j,\rmI}(a,a')$, for which, we give a numeric example in Section S7.1 of the Online Supplement. In the special case where $w_{ijk} = N_{ij}^{-1}$, i.e., $\pi_{ijk}=(IN_{ij})^{-1}$, and $\bfX_{ijk}$ does not vary with $k$, $\bfgamma_{j,\rmI}(a)$ equals the coefficient of $\widetilde\bfX_{ij\cdot}^c$ in the OLS fit of $\widetilde\epsilon_{ij\cdot}(a)$ on $\widetilde\bfX_{ij\cdot}^c$, and therefore $\widehat\tau_{j,\rmI}^\adj(a,a')$ has a smaller asymptotic variance than $\widehat\tau_{j,\rmI}(a,a')$. 
	
	Alternatively, we can consider the following analysis of covariance (ANCOVA) regression model adapted from \citet{Schochet2021} for a blocked CRE:
	\begin{equation} \label{eq:model-I-ancova}
		Y_{ijk} = \suma\beta_j(a)G_i(a)+\bfX_{ijk}^c\bfgamma_j+e_{ijk},
	\end{equation}
	with a shared covariate vector $\bfgamma_j$ across different treatment adoption times. A regression estimator for $\tau_j(a,a')$ can be obtained via the WLS fit of \eqref{eq:model-I-ancova} with individual weight $\pi_{ijk}$, which is in the form of $\widehat\tau_{j,\rmI}^\ancova(a,a')=\widehat\beta_{j,\rmI}^\ancova(a)-\widehat\beta_{j,\rmI}^\ancova(a')$, with $\widehat\beta_{j,\rmI}^\ancova(a)=\overline y_{\cdot j\cdot}(a)-\overline\bfx_{\cdot j\cdot}^c(a)\widehat\bfgamma_{j,\rmI}$. To obtain $\widehat\Cov_\CR\{\widehat\bftau_{\rmI}^\ancova(a,a')\}$, we have
	\begin{align*}
		\bfE_{\rmI}^\ancova&=\left(\sum_{i=1}^I(\bfD_{i,\rmI}^\ancova)^\top\bfPi_{i,\rmI}\bfD_{i,\rmI}^\ancova\right)^{-1}\left(\sumi(\bfD_{i,\rmI}^\ancova)^\top\bfPi_{i,\rmI}\widehat{\bfe}_{i,\rmI}^\ancova(\widehat{\bfe}_{i,\rmI}^\ancova)^\top\bfPi_{i,\rmI}\bfD_{i,\rmI}^\ancova\right)\displaybreak[0]\\
		&\qquad\times\left(\sum_{i=1}^I(\bfD_{i,\rmI}^\ancova)^\top\bfPi_{i,\rmI}\bfD_{i,\rmI}^\ancova\right)^{-1},
	\end{align*}
	where $\widehat\bfe_{i,\rmI}^\ancova$ is the vector of residuals from the WLS fit of \eqref{eq:model-I-ancova}, $\bfD_{i,\rmI}^\ancova$ is of dimensions $N_{i\cdot}\times J(J+1+p_x)$ with $\bfD_{i,\rmI}^\ancova=((\bfD_{i1,\rmI}^\ancova)^\top\cdots (\bfD_{iJ,\rmI}^\ancova)^\top)^\top$ and the $k$-th row of $\bfD_{ij,\rmI}^\ancova$ being $(\bfG_i(\calA)^\top\otimes\pmb\bbone_{ijk}^\top,\pmb\bbone_{ijk}^\top\otimes\bfX_{ijk}^c)$. The CR variance estimator for $\Cov\{\widehat\bftau_{\rmI}^\ancova(a,a')\}$ is $\widehat\Cov_\CR\{\widehat\bftau_{\rmI}^\ancova(a,a')\}=\bfQ(a,a')\bfE_{\rmI,[1\rightarrow J(J+1),1\rightarrow J(J+1)]}^\ancova\bfQ(a,a')^\top$. Define $q(a)=I(a)/I$ as $I\rightarrow\infty$ for $a\in\calA$. The theoretical properties of $\widehat\bftau_{\rmI}^\ancova(a,a')$ and $\widehat\Cov_\CR\{\widehat\bftau_{\rmI}^\ancova(a,a')\}$ are given in the following Corollary \ref{coro:consistency-AN-I-ancova}.
	
	\begin{corollary} \label{coro:consistency-AN-I-ancova}
		Define $r_{ij\cdot}(a)=\widetilde\epsilon_{ij\cdot}(a)-\widetilde\bfX_{ij\cdot}^c\suma q(a)\bfgamma_{j,\rmI}(a)$ and $\bfr_{i\cdot}(a)=(r_{i1\cdot}(a),\allowbreak\ldots,r_{iJ\cdot}(a))^\top$. Under Assumptions \ref{asp:sutva} and \ref{asp:rand}, and regularity conditions (C1) - (C3) in Section S2 of the Online Supplement, for $a<a'\in\calA$, $\widehat\bftau_{\rmI}^\ancova(a,a')=\bftau(a,a')+\bfo_\bbP(1)$; if further $\max_i\pi_{ij\cdot}=o(I^{-2/3})$ for $j=1,\ldots,J$ and $\bfV(\bfr_{i\cdot})(a,a')\nrightarrow \bm{0}_{J\times J}$, then $I^{1/2}\{\widehat\bftau_{\rmI}^\ancova(a,a')-\bftau(a,a')\}\allowbreak\xrightarrow{d}\calN(\bm{0}_J,\bfV(\bfr_{i\cdot})(a,a'))$, and $I\times\widehat \Cov_\CR\{\widehat\bftau_{\rmI}^\ancova(a,a')\}=\bfV_c(\bfr_{i\cdot})(a,a')+\bfo_\bbP(1)$.
	\end{corollary}
	
	Corollary \ref{coro:consistency-AN-I-ancova} presents the consistency and asymptotic normality of $\widehat\bftau_{\rmI}^\ancova(a,a')$, and the conservativeness of $\widehat\Cov_\CR\{\widehat\bftau_{\rmI}^\ancova(a,a')\}$, in the L\"{o}wner ordering. As suggested in \citet{Su2021}, each scalar treatment effect estimator $\widehat\tau_{j,\rmI}^\ancova(a,a')$ can be less efficient than $\widehat\tau_{j,\rmI}(a,a')$, and $\widehat\tau_{j,\rmI}^\adj(a,a')$ does not necessarily improve $\widehat\tau_{j,\rmI}^\ancova(a,a')$ in terms of asymptotic efficiency. Thus, we do not recommend $\widehat\tau_{j,\rmI}^\ancova(a,a')$ because it is suboptimal even in simpler cases \citep{Lin2013}, unless it is expected that there is little treatment heterogeneity by $\bfX_{ijk}$.
	
	\subsection{Estimators using cluster-level data} \label{sec:est-clu}
	
	\subsubsection{Using cluster-period averages}
	
	Motivated by the expression in \eqref{eq:d-wate-A}, $\tau_j(a,a')$ can be estimated using cluster-period average $\overline Y_{ij\cdot}$ and cluster weight $\pi_{ij\cdot}$, and the setting can be viewed as an SR-IRE with $\overline Y_{ij\cdot}$ as the observed outcome \citep{Roth2023}. In this case, we consider the cluster-period mean regression model
	\begin{equation}  \label{eq:model-A}
		\overline Y_{ij\cdot} = \suma\beta_j(a)G_i(a)+e_{ij},
	\end{equation}
	where $e_{ij}$ is the cluster-level random noise assumed to have a mean-zero distribution and is independent across clusters. We can obtain an estimator for $\tau_j(a,a')$ by fitting \eqref{eq:model-A} with cluster weight $\pi_{ij\cdot}$, which is in the form of $\widehat\tau_{j,\rmA}(a,a')=\widehat\beta_{j,\rmA}(a)-\widehat\beta_{j,\rmA}(a')$, with $\widehat\beta_{j,\rmA}(a)=\overline y_{\cdot j\cdot}(a)$. This estimator is equivalent to $\widehat\tau_{j,\rmI}(a,a')$. Here, we use the subscript `$\rmA$' to clarify that the estimator is obtained using cluster-period averages. 
	
	The covariance matrix $\Cov\{\widehat\bftau_{\rmA}(a,a')\}$ can be estimated by obtaining the HC variance estimator for the vector of estimated regression coefficients,
	\begin{align*} %\label{eq:CRSE-beta-I}
		\bfE_{\rmA}=\left(\sum_{i=1}^I\bfD_{i,\rmA}^\top\bfPi_{i,\rmA}\bfD_{i,\rmA}\right)^{-1}\left(\sumi\bfD_{i,\rmA}^\top\bfPi_{i,\rmA}\widehat{\bfe}_{i,\rmA}\widehat{\bfe}_{i,\rmA}^\top\bfPi_{i,\rmA}\bfD_{i,\rmA}\right)\left(\sum_{i=1}^I\bfD_{i,\rmA}^\top\bfPi_{i,\rmA}\bfD_{i,\rmA}\right)^{-1},
	\end{align*}
	where $\widehat\bfe_{i,\rmA}$ is the $J$-dimensional residual vector from the WLS fit of \eqref{eq:model-A}, $\bfD_{i,\rmA}$ is of dimensions $J\times J(J+1)$ with the $j$-th row being $\bfG_i(\calA)^\top\otimes\pmb\bbone_{ij}^\top$, and $\bfPi_{i,\rmA}=\diag(\pi_{i1\cdot},\ldots,\pi_{iJ\cdot})$. Similar to $\pmb\bbone_{ijk}$, here, $\pmb\bbone_{ij}=(\bbone_{ij,1}\ldots,\bbone_{ij,J})^\top$ with $\bbone_{ij,j'}=1$ if $j=j'$ and $\bbone_{ij,j'}=0$ otherwise for $j'=1,\ldots,J$. Then, the HC variance estimator for $\Cov\{\widehat\bftau_{\rmA}(a,a')\}$ is $\widehat\Cov_\HC\{\widehat\bftau_{\rmA}(a,a')\}=\bfQ(a,a')\bfE_\rmA\bfQ(a,a')^\top$, and the HC variance estimator for $\se^2\{\widehat\tau_{j,\rmA}(a,a')\}$ is
	\begin{align*} %\label{eq:HW-A}
		\widehat \se_\HC^2\{\widehat\tau_{j,\rmA}(a,a')\}&=\widehat\Cov_\HC\{\widehat\bftau_{\rmA}(a,a')\}_{(j,j)}
		=\sumi\frac{G_i(a)}{\pi_{\cdot j\cdot}(a)^2}\left(\pi_{ij\cdot}\widehat e_{ij}\right)^2+\sumi\frac{G_i(a')}{\pi_{\cdot j\cdot}(a')^2}\left(\pi_{ij\cdot}\widehat e_{ij}\right)^2.
	\end{align*}
	Importantly, the HC variance estimator is a special case of the CR variance estimator with cluster size $N_{ij}=1$ for all $i=1,\ldots,I$ and $j=1,\ldots,J$. The consistency and asymptotic normality of $\widehat\bftau_{\rmA}(a,a')$ as well as the conservativeness of $\widehat\Cov_\HC\{\widehat\bftau_{\rmA}(a,a')\}$, in the L\"{o}wner ordering, are given in Theorem \ref{thm:consistency-AN-A}.
	
	\begin{theorem} \label{thm:consistency-AN-A}
		Under Assumptions \ref{asp:sutva} and \ref{asp:rand}, and regularity conditions (C1) and (C4) in Section S2 of the Online Supplement, for $a<a'\in\calA$, $\widehat\bftau_{\rmA}(a,a')=\bftau_j(a,a')+\bfo_\bbP(1)$; if further $\max_i\pi_{ij\cdot}=o(I^{-2/3})$ for $j=1,\ldots,J$ and $\bfV(\widetilde\bfepsilon_{i\cdot})(a,a')\nrightarrow \bm{0}_{J\times J}$, then $I^{1/2}\{\widehat\bftau_{\rmA}(a,a')-\bftau(a,a')\}\xrightarrow{d}\calN(\bm{0}_J,\bfV(\widetilde\bfepsilon_{i\cdot})(a,a'))$, and $I\times\widehat \Cov_\HC\{\widehat\bftau_{\rmA}(a,a')\}=\bfV_c(\widetilde\bfepsilon_{i\cdot})(a,a')+\bfo_\bbP(1)$.
	\end{theorem}
	
	Similar to the estimators using individual-level data, covariate adjustment can be incorporated into \eqref{eq:model-A} to obtain the covariate-adjusted cluster-period mean regression model:
	\begin{equation}  \label{eq:model-A-cov}
		\overline Y_{ij\cdot} = \suma\beta_j(a)G_i(a)+\suma G_i(a)\bfC_{ij}^c\bfgamma_j(a)+e_{ij},
	\end{equation}
	where the centered cluster-level covariate vector $\bfC_{ij}^c=\bfC_{ij}-\overline\bfC_{\cdot j}$ with $\overline\bfC_{\cdot j}=\sumi \pi_{ij\cdot}\bfC_{ij}$. With cluster weight $\pi_{ij\cdot}$, we can obtain estimates $\widehat\beta_{j,\rmA}^\adj(a)$ and $\widehat\bfgamma_{j,\rmA}(a)$ from the WLS fit of \eqref{eq:model-A-cov}, which leads to a covariate-adjusted estimator for $\tau_j(a,a')$ in the form of $\widehat\tau_{j,\rmA}^\adj(a,a')=\widehat\beta_{j,\rmA}^\adj(a)-\widehat\beta_{j,\rmA}^\adj(a')$, with $\widehat\beta_{j,\rmA}^\adj(a)=\overline y_{\cdot j\cdot}(a)-\overline\bfc_{\cdot j}^c(a)\widehat\bfgamma_{j,\rmA}(a)$ where $\overline\bfc_{\cdot j}^c(a)=\sumi \pi_{ij\cdot}G_i(a)\bfC_{ij}^c/\pi_{\cdot j\cdot}(a)$. To obtain $\widehat \Cov_\HC\{\widehat\bftau_{\rmA}^\adj(a,a')\}$, we have
	\begin{align*} %\label{eq:CRSE-beta-I}
		\bfE_{\rmA}^\adj&=\left(\sum_{i=1}^I(\bfD_{i,\rmA}^\adj)^\top\bfPi_{i,\rmA}\bfD_{i,\rmA}^\adj\right)^{-1}\left(\sumi(\bfD_{i,\rmA}^\adj)^\top\bfPi_{i,\rmA}\widehat{\bfe}_{i,\rmA}^\adj(\widehat{\bfe}_{i,\rmA}^\adj)^\top\bfPi_{i,\rmA}\bfD_{i,\rmA}^\adj\right)\displaybreak[0]\\
		&\qquad\times\left(\sum_{i=1}^I(\bfD_{i,\rmA}^\adj)^\top\bfPi_{i,\rmA}\bfD_{i,\rmA}^\adj\right)^{-1},
	\end{align*}
	where $\widehat\bfe_{i,\rmA}^\adj$ is the $J$-dimensional vector of residuals from the WLS fit of \eqref{eq:model-A-cov}, $\bfD_{i,\rmA}^\adj$ is of dimensions $J\times J(1+p_c)(J+1)$ with the $j$-th row being $(\bfG_i(\calA)^\top\otimes\pmb\bbone_{ij}^\top,\bfG_i(\calA)^\top\otimes(\pmb\bbone_{ij}^\top\otimes\bfC_{ij}^c))$. The HC variance estimator for $\Cov\{\widehat\bftau_{\rmA}^\adj(a,a')\}$ is $\widehat\Cov_\HC\{\widehat\bftau_{\rmA}^\adj(a,a')\}=\bfQ(a,a')\bfE_{\rmA,[1\rightarrow J(J+1),1\rightarrow J(J+1)]}^\adj\times\allowbreak\bfQ(a,a')^\top$. Let $\bfgamma_{j,\rmA}(a)$ denote the WLS coefficient vectors obtained using cluster-period averages if the full set of potential outcomes is available. That is, regress $\overline Y_{ij\cdot}(a)$ on $(1,\bfC_{ij}^c)$ with weight $\pi_{ij\cdot}$ assuming $\overline Y_{ij\cdot}(a)$ is available for each cluster $i$ in period $j$. The theoretical properties of $\widehat\bftau_{\rmA}^\adj(a,a')$ and $\widehat\Cov_\HC\{\widehat\bftau_{\rmA}^\adj(a,a')\}$ are given in Theorem \ref{thm:consistency-AN-A-cov}.
	\begin{theorem} \label{thm:consistency-AN-A-cov}
		Define $r_{ij\cdot}(a)=\widetilde\epsilon_{ij\cdot}(a)-I\pi_{ij\cdot}\bfC_{ij}^c\bfgamma_{j,\rmA}(a)$ and $\bfr_{i\cdot}(a)=(r_{i1\cdot}(a),\ldots,\allowbreak r_{iJ\cdot}(a))^\top$. Under Assumptions \ref{asp:sutva} and \ref{asp:rand}, and regularity conditions (C1), (C4), and (C5) in Section S2 of the Online Supplement, for $a<a'\in\calA$, $\widehat\bftau_{\rmA}^\adj(a,a')=\bftau(a,a')+\bfo_\bbP(1)$; if further $\max_i\pi_{ij\cdot}=o(I^{-2/3})$ and $\bfV(\bfr_{i\cdot})(a,a')\nrightarrow \bm{0}_{J\times J}$, then $I^{1/2}\{\widehat\bftau_{\rmA}^\adj(a,a')-\bftau(a,a')\}\xrightarrow{d}\calN(\bm{0}_J,\bfV(\bfr_{i\cdot})(a,a'))$, and $I\times\widehat \Cov_\HC\{\widehat\bftau_{\rmA}^\adj(a,a')\}\allowbreak=\bfV_c(\bfr_{i\cdot})(a,a')+\bfo_\bbP(1)$.
	\end{theorem}
	
	Theorem \ref{thm:consistency-AN-A-cov} presents the consistency and asymptotic normality of $\widehat\bftau_{\rmA}^\adj(a,a')$, and the conservativeness of $\widehat\Cov_\HC\{\widehat\bftau_{\rmA}^\adj(a,a')\}$, in the L\"{o}wner ordering. For the variance $\se^2\{\widehat\tau_{j,\rmA}^\adj(a,a')\}$, the residual $r_{ij\cdot}(a)$ in Theorem \ref{thm:consistency-AN-A-cov} is in the form of $\widetilde\epsilon_{ij\cdot}(a)-I\pi_{ij\cdot}\bfC_{ij}^c\bfgamma$, which suggests that the optimal choice of $\bfgamma$ is the coefficient of $I\pi_{ij\cdot}\bfC_{ij}^c$ in the OLS fit of $\widetilde\epsilon_{ij\cdot}(a)$ on $I\pi_{ij\cdot}\bfC_{ij}^c$, rather than $\bfgamma_{j,\rmA}(a)$ from the WLS fit using cluster averages. Similar to the discussion following Theorem \ref{thm:consistency-AN-I-cov}, $\widehat\tau_{j,\rmA}^\adj(a,a')$ may not lead to better asymptotic efficiency compared to $\widehat\tau_{j,\rmA}(a,a')$. We omit the detailed discussion of $\widehat\tau_{j,\rmA}^\ancova(a,a')$, the ANCOVA estimator obtained via regressing $\overline Y_{ij\cdot}$ on $(\bfG_i(\calA)^\top\otimes\pmb\bbone_{ij}^\top,\pmb\bbone_{ij}^\top\otimes\bfC_{ij}^c)$ with weights $\pi_{ij\cdot}$, because it is suboptimal and also may not improve the asymptotic efficiency of $\widehat\tau_{j,\rmA}(a,a')$. Similar to $\widehat\bftau_{\rmA}(a,a')$ and $\widehat\bftau_{\rmA}^\adj(a,a')$, $\widehat\bftau_{\rmA}^\ancova(a,a')$ is consistent and asymptotically normal for $\bftau(a,a')$, and the associated HC variance estimator $\widehat\Cov_\HC\{\widehat\bftau_{\rmA}^\ancova(a,a')\}$ is conservative for $\Cov\{\widehat\bftau_{\rmA}^\ancova(a,a')\}$. We provide some clarifications on $\widehat\tau_{j,\rmA}(a,a')$ and $\widehat\tau_{j,\rmA}^\adj(a,a')$ regarding the special cases of $w_{ijk}=N_{ij}^{-1}$ and $w_{ijk}=1$ in Remark S3 in Section S2 of the Online Supplement.

	\subsubsection{Using scaled cluster-period totals}
	
	The expression in \eqref{eq:d-wate-T} suggests that $\tau_j(a,a')$ can be estimated using scaled cluster-period totals $\widetilde Y_{ij\cdot}$. Similar to estimating $\tau_j(a,a')$ using cluster-period averages, using the scaled cluster-period totals also reduces to an SR-IRE with $\widetilde Y_{ij\cdot}$ as observed outcomes. Consider the following scaled cluster-period total regression model
	\begin{equation}  \label{eq:model-T}
		\widetilde Y_{ij\cdot} = \suma\beta_j(a)G_i(a)+e_{ij}.
	\end{equation}
	We can obtain an estimator for $\tau_j(a,a')$ via the OLS fit of \eqref{eq:model-T}, which is in the form of $\widehat\tau_{j,\rmT}(a,a')=\widehat\beta_{j,\rmT}(a)-\widehat\beta_{j,\rmT}(a')$, with $\widehat\beta_{j,\rmT}(a)=\widetilde y_{\cdot j\cdot}(a)=\sumi G_i(a)\widetilde Y_{ij\cdot}(a)/\allowbreak\sumi G_i(a)$. We use the subscript `$\rmT$' to clarify that the estimator is obtained using scaled cluster-period totals. The covariance matrix $\Cov\{\widehat\bftau_{\rmT}(a,a')\}$ can be estimated by the HC variance estimator. Specifically, we have
	\begin{align*} %\label{eq:CRSE-beta-I}
		\bfE_{\rmT}=\left(\sum_{i=1}^I\bfD_{i,\rmT}^\top\bfPi_{i,\rmT}\bfD_{i,\rmT}\right)^{-1}\left(\sumi\bfD_{i,\rmT}^\top\bfPi_{i,\rmT}\widehat{\bfe}_{i,\rmT}\widehat{\bfe}_{i,\rmT}^\top\bfPi_{i,\rmT}\bfD_{i,\rmT}\right)\left(\sum_{i=1}^I\bfD_{i,\rmT}^\top\bfPi_{i,\rmT}\bfD_{i,\rmT}\right)^{-1},
	\end{align*}
	where $\widehat\bfe_{i,\rmT}$ is the $J$-dimensional vector of residuals from the OLS fit of \eqref{eq:model-T}, $\bfD_{i,\rmT}=\bfD_{i,\rmA}$, and $\bfPi_{i,\rmT}=\bfI_{J\times J}$. The HC variance estimator for $\Cov\{\widehat\bftau_{\rmT}(a,a')\}$ is $\widehat\Cov_\HC\{\widehat\bftau_{\rmT}(a,a')\}=\bfQ(a,a')\bfE_\rmT\bfQ(a,a')^\top$. The consistency and asymptotic normality of $\widehat\bftau_{\rmT}(a,a')$ and the conservativeness of $\widehat\Cov_\HC\{\widehat\bftau_{\rmT}(a,a')\}$ are given in Theorem \ref{thm:consistency-AN-T}.
	
	\begin{theorem} \label{thm:consistency-AN-T}
		Define $r_{ij\cdot}(a)=\widetilde Y_{ij\cdot}(a)-\overline Y_{\cdot j\cdot}(a)$ and $\bfr_{i\cdot}(a)=(r_{i1\cdot}(a),\ldots,r_{iJ\cdot}(a))^\top$. Under Assumptions \ref{asp:sutva} and \ref{asp:rand}, and regularity conditions (C1) and (C6) in Section S2 of the Online Supplement, for $a<a'\in\calA$, $\widehat\bftau_{\rmT}(a,a')=\bftau(a,a')+\bfo_\bbP(1)$; if further $I^{-2}\sumi\widetilde Y_{ij\cdot}(a)^4=o(1)$ for $j=1,\ldots,J$ and $\bfV(\bfr_{i\cdot})(a,a')\nrightarrow \bm{0}_{J\times J}$, then $I^{1/2}\{\widehat\bftau_{\rmT}(a,a')-\bftau(a,a')\}\xrightarrow{d}\calN(\bm{0}_J,\bfV(\bfr_{i\cdot})(a,a'))$, and $I\times\widehat \Cov_\HC\{\widehat\bftau_{\rmT}(a,a')\}=\bfV_c(\bfr_{i\cdot})(a,a')+\bfo_\bbP(1)$.
	\end{theorem}
	
	Similarly, covariate adjustment can be incorporated into \eqref{eq:model-T} to obtain the following fully-interacted regression model:
	\begin{equation}  \label{eq:model-T-cov}
		\widetilde Y_{ij\cdot} = \suma\beta_j(a)G_i(a)+\suma G_i(a)\bfC_{ij}^c\bfgamma_j(a)+e_{ij},
	\end{equation}
	where $\bfC_{ij}^c=\bfC_{ij}-\overline\bfC_{\cdot j}$ with $\overline\bfC_{\cdot j}=I^{-1}\sumi\bfC_{ij}$. We can obtain estimates $\widehat\beta_{j,\rmT}^\adj(a)$ and $\widehat\bfgamma_{j,\rmT}(a)$ from the OLS fit of \eqref{eq:model-T-cov}, leading to a covariate-adjusted estimator for $\tau_j(a,a')$ in the form of $\widehat\tau_{j,\rmT}^\adj(a,a')=\widehat\beta_{j,\rmT}^\adj(a)-\widehat\beta_{j,\rmT}^\adj(a')$, with $\widehat\beta_{j,\rmT}^\adj(a)=\widetilde y_{\cdot j\cdot}(a)-\overline\bfc_{\cdot j,\rmT}^c(a)\widehat\bfgamma_{j,\rmT}(a)$ where $\overline\bfc_{\cdot j,\rmT}^c(a)=\sumi G_i(a)\bfC_{ij}^c/\allowbreak\sumi G_i(a)$. To obtain $\widehat \Cov_\HC\{\widehat\bftau_{\rmT}^\adj(a,a')\}$, we have
	\begin{align*} %\label{eq:CRSE-beta-I}
		\bfE_{\rmT}^\adj&=\left(\sum_{i=1}^I(\bfD_{i,\rmT}^\adj)^\top\bfPi_{i,\rmT}\bfD_{i,\rmT}^\adj\right)^{-1}\left(\sumi(\bfD_{i,\rmT}^\adj)^\top\bfPi_{i,\rmT}\widehat{\bfe}_{i,\rmT}^\adj(\widehat{\bfe}_{i,\rmT}^\adj)^\top\bfPi_{i,\rmT}\bfD_{i,\rmT}^\adj\right)\displaybreak[0]\\
		&\qquad\times\left(\sum_{i=1}^I(\bfD_{i,\rmT}^\adj)^\top\bfPi_{i,\rmT}\bfD_{i,\rmT}^\adj\right)^{-1},
	\end{align*}
	where $\widehat\bfe_{i,\rmT}^\adj$ is the $J$-dimensional vector of residuals from the OLS fit of \eqref{eq:model-T-cov}, and $\bfD_{i,\rmT}^\adj=\bfD_{i,\rmA}^\adj$. The HC variance estimator for $\Cov\{\widehat\bftau_{\rmT}^\adj(a,a')\}$ is $\widehat\Cov_\HC\{\widehat\bftau_{\rmT}^\adj(a,a')\}=\bfQ(a,a')\times\allowbreak\bfE_{\rmT,[1\rightarrow J(J+1),1\rightarrow J(J+1)]}^\adj\bfQ(a,a')^\top$. Let $\bfgamma_{j,\rmT}(a)$ denote the OLS coefficient vectors obtained using scaled cluster-period totals if the full set of potential outcomes is available. That is, regress $\widetilde Y_{ij\cdot}(a)$ on $(1,\bfC_{ij}^c)$ assuming $\widetilde Y_{ij\cdot}(a)$ is available for each cluster $i$ in period $j$. The consistency and asymptotic normality of $\widehat\bftau_{\rmT}^\adj(a,a')$ and the conservativeness of $\widehat\Cov_\HC\{\widehat\bftau_{\rmT}^\adj(a,a')\}$, in the L\"{o}wner ordering, are given in Theorem \ref{thm:consistency-AN-T-cov}.
	\begin{theorem} \label{thm:consistency-AN-T-cov}
		Define $r_{ij\cdot}(a)=\widetilde Y_{ij\cdot}(a)-\overline Y_{\cdot j\cdot}(a)-\bfC_{ij}^c\bfgamma_{j,\rmT}(a)$ and $\bfr_{i\cdot}(a)=(r_{i1\cdot}(a),\ldots,\allowbreak r_{iJ\cdot}(a))^\top$. Under Assumptions \ref{asp:sutva} and \ref{asp:rand}, and regularity conditions (C1), (C6), and (C7) in Section S2 of the Online Supplement, for $a<a'\in\calA$, $\widehat\bftau_{\rmT}^\adj(a,a')=\bftau(a,a')+\bfo_\bbP(1)$; if further $I^{-1}\sum_i\widetilde Y_{ij\cdot}(a)^4=O(1)$ for $j=1,\ldots,J$ and $\bfV(\bfr_{i\cdot})(a,a')\nrightarrow \bm{0}_{J\times J}$, then $I^{1/2}\{\widehat\bftau_{\rmT}^\adj(a,a')-\bftau(a,a')\}\xrightarrow{d}\calN(\bm{0}_J,\bfV(\bfr_{i\cdot})(a,a'))$, and $I\times\widehat \Cov_\HC\{\widehat\bftau_{\rmT}^\adj(a,a')\}=\bfV_c(\bfr_{i\cdot})(a,a')+\bfo_\bbP(1)$.
	\end{theorem}
	
	Comparing results in Theorems \ref{thm:consistency-AN-T} and \ref{thm:consistency-AN-T-cov}, for marginal variances, we find that covariate adjustment always improves the asymptotic efficiency because $\se^2\{\widehat\tau_{j,\rmT}^\adj(a,a')\}\leq \se^2\{\widehat\tau_{j,\rmT}(a,a')\}$ from the regressions using scaled cluster-period totals. Theorems \ref{thm:consistency-AN-T} and \ref{thm:consistency-AN-T-cov} also provide the theoretical foundation for using the HC variance estimator for variance estimation in SR-IREs, which was pointed out in \citet{Roth2023} without a theoretical proof. Finally, the ANCOVA estimator $\widehat\tau_{j,\rmT}^\ancova(a,a')$ obtained via the OLS fit of $\widetilde Y_{ij\cdot}$ on $(\bfG_i(\calA)^\top\otimes\pmb\bbone_{ij}^\top,\pmb\bbone_{ij}^\top\otimes\bfC_{ij}^c)$ is less efficient than $\widehat\tau_{j,\rmT}^\adj(a,a')$ and may not improve the asymptotic efficiency of $\widehat\tau_{j,\rmT}(a,a')$, as in simpler cases under parallel-arm CREs \citep{Su2021}. Similar $\widehat\bftau_{\rmI}^\ancova(a,a')$ and $\widehat\bftau_{\rmA}^\ancova(a,a')$, $\widehat\bftau_{\rmT}^\ancova(a,a')$ is also consistent and asymptotically normal for $\bftau(a,a')$, and the associated HC variance estimator $\widehat\Cov_\HC\{\widehat\bftau_{\rmT}^\ancova(a,a')\}$ is conservative for $\Cov\{\widehat\bftau_{\rmT}^\ancova(a,a')\}$, in the L\"{o}wner ordering.
	
	\begin{remark} \label{rmk:rmk-4}
		The following results hold when choosing $w_{ijk}=N_{ij}^{-1}$ to target the cluster-average estimands. Without covariate adjustment, we have $\widehat\bftau_{\rmI}(a,a')=\widehat\bftau_{\rmA}(a,a')=\widehat\bftau_{\rmT}(a,a')$, and also $\widehat\Cov_\CR\{\widehat\bftau_{\rmI}(a,a')\}=\widehat\Cov_\HC\{\widehat\bftau_{\rmA}(a,a')\}=\widehat\Cov_\HC\{\widehat\bftau_{\rmT}(a,a')\}$. With covariate adjustment, we have $\widehat\bftau_{\rmA}^\adj(a,a')=\widehat\bftau_{\rmT}^\adj(a,a')$ and $\widehat\Cov_\HC\{\widehat\bftau_{\rmA}^\adj(a,a')\}=\widehat\Cov_\HC\{\widehat\bftau_{\rmT}^\adj(a,a')\}$; if we only adjust for cluster-level covariates in $\widehat\bftau_{\rmI}^\adj(a,a')$, then $\widehat\bftau_{\rmI}^\adj(a,a')=\widehat\bftau_{\rmA}^\adj(a,a')=\widehat\bftau_{\rmT}^\adj(a,a')$ and $\widehat\Cov_\CR\{\widehat\bftau_{\rmI}^\adj(a,a')\}=\widehat\Cov_\HC\{\widehat\bftau_{\rmA}^\adj(a,a')\}=\widehat\Cov_\HC\{\widehat\bftau_{\rmT}^\adj(a,a')\}$.
	\end{remark}
	
	\subsection{Generalization to summary estimands}
	
	The summary estimand in \eqref{eq:gen-estimand} is a weighted sum of $\tau_j(a,a')$ for $j=1,\ldots,J$, assuming $a<a'\in\calA$. We can estimate $\theta$ via plugging in estimates of $\tau_j(a,a')$ from any proposed regression estimators, e.g., $\widehat\theta_{\rmI}^\adj = \sumj\sum_{a<a'}b_{\theta,j}(a,a')\widehat\tau_{j,\rmI}^\adj(a,a')$ if we select the regression estimator using individual-level data with covariate adjustment. The variance of $\widehat\theta$, $\se^2(\widehat\theta)=\bfb_\theta^\top\Cov(\widehat\bftau)\bfb_\theta$, where $\bfb_\theta=(\bfb_\theta(1,2),\ldots,\bfb_\theta(1,\infty)^\top,\bfb_\theta(2,3)^\top,\ldots,\allowbreak\bfb_\theta(2,\infty)^\top,\ldots,\allowbreak\bfb_\theta(J,\infty)^\top)^\top$ with $\bfb_\theta(a,a')=(b_{\theta,1}(a,a'),\allowbreak\ldots,b_{\theta,j}(a,a'))^\top$ for $a,a\in\calA$ and $a<a'$, can be estimated using the CR or HC variance estimator depending on the selected regression estimator. Continuing with the example of $\widehat\theta_{\rmI}^\adj$, we have the CR variance estimator as $\widehat\se_\CR^2(\widehat\theta_{\rmI}^\adj)=\bfb_\theta^\top\widehat\Cov_\CR(\widehat\bftau_\rmI^\adj)\bfb_\theta$, where $\widehat\Cov_\CR(\widehat\bftau_\rmI^\adj)=\bfQ\bfE_{\rmI,[1\rightarrow J(J+1),1\rightarrow J(J+1)]}^\adj\bfQ^\top$ with
	\begin{align*}
		\bfQ=\begin{pmatrix}
			\bfQ(1,2)^\top & \cdots & \bfQ(1,\infty)^\top & \bfQ(2,3)^\top & \cdots & \bfQ(2,\infty)^\top & \cdots & \bfQ(J,\infty)^\top
		\end{pmatrix}^\top.
	\end{align*}
	Table \ref{tab:reg-est-cov-est} summarizes regression estimators and their corresponding covariance estimators. For conciseness, we give the consistency and asymptotic normality of $\widehat\theta$, and the conservativeness of $\widehat\se^2(\widehat\theta)$ in Theorem S1 in Section S3 of the Online Supplement.
	\begin{table}[htbp] %***
		\caption{Regression estimators and their corresponding covariance estimators.}\label{tab:reg-est-cov-est}
		\vspace{-0.15in}
		\begin{center}
			\resizebox{\linewidth}{!}{
				\begin{tabular}{llll} 
					\hline %***5truept
					\textbf{Estimator} & \textbf{Covariance estimator} & \textbf{Estimator} & \textbf{Covariance estimator} \\
					\hline
					$\widehat\bftau_\rmI$ & $\widehat\Cov_\CR(\widehat\bftau_\rmI)=\bfQ\bfE_\rmI\bfQ^\top$ & $\widehat\bftau_\rmA$ & $\widehat\Cov_\HC(\widehat\bftau_\rmA)=\bfQ\bfE_\rmA\bfQ^\top$ \\
					$\widehat\bftau_\rmI^\adj$ & $\widehat\Cov_\CR(\widehat\bftau_\rmI^\adj)=\bfQ\bfE_{\rmI,[1\rightarrow J(J+1),1\rightarrow J(J+1)]}^\adj\bfQ^\top$ & $\widehat\bftau_\rmA^\adj$ & $\widehat\Cov_\HC(\widehat\bftau_\rmA)=\bfQ\bfE_{\rmA,[1\rightarrow J(J+1),1\rightarrow J(J+1)]}^\adj\bfQ^\top$ \\
					$\widehat\bftau_\rmI^\ancova$ & $\widehat\Cov_\CR(\widehat\bftau_\rmI^\ancova)=\bfQ\bfE_{\rmI,[1\rightarrow J(J+1),1\rightarrow J(J+1)]}^\ancova\bfQ^\top$ & $\widehat\bftau_\rmT$ & $\widehat\Cov_\HC(\widehat\bftau_\rmT)=\bfQ\bfE_\rmT\bfQ^\top$ \\
					&& $\widehat\bftau_\rmT^\adj$ & $\widehat\Cov_\HC(\widehat\bftau_\rmT)=\bfQ\bfE_{\rmT,[1\rightarrow J(J+1),1\rightarrow J(J+1)]}^\adj\bfQ^\top$ \\
					\hline
				\end{tabular}
			}
		\end{center}
	\end{table}
	
	\section{Comparisons and recommendations}  \label{sec:comp}
	
	We compare different regression estimators in terms of their efficiency in estimating $\tau_j(a,a')$ and $\theta$, and provide relevant recommendations. For ease of exposition, we first focus on the individual DWATE estimand before returning to the summary estimands in Section \ref{sec:gen_compare}. The theoretical results of regression estimators, marginal standard errors, and corresponding theorems are summarized in Table \ref{tab:results-summary}.
	\begin{table}[htbp] %***
		\caption{Estimators, standard errors, and corresponding theoretical results.}\label{tab:results-summary}
		\vspace{-0.15in}
		\begin{center}
			\resizebox{\linewidth}{!}{
				\begin{tabular}{llllll} 
					\hline %***5truept
					\textbf{Estimator} & \textbf{Standard error} & \textbf{Result} & \textbf{Estimator} & \textbf{Standard error} & \textbf{Result} \\
					\hline
					$\widehat\tau_{j,\rmI}(a,a')$ & $\widehat\se_\CR\{\widehat\tau_{j,\rmI}(a,a')\}$ & Theorem \ref{thm:consistency-AN-I} & $\widehat\tau_{j,\rmA}(a,a')$ & $\widehat\se_\HC\{\widehat\tau_{j,\rmA}(a,a')\}$ & Theorem \ref{thm:consistency-AN-A} \\
					$\widehat\tau_{j,\rmI}^\adj(a,a')$ & $\widehat\se_\CR\{\widehat\tau_{j,\rmI}^\adj(a,a')\}$ & Theorem \ref{thm:consistency-AN-I-cov} & $\widehat\tau_{j,\rmA}^\adj(a,a')$ & $\widehat\se_\HC\{\widehat\tau_{j,\rmA}^\adj(a,a')\}$ & Theorem \ref{thm:consistency-AN-A-cov} \\
					$\widehat\tau_{j,\rmI}^\ancova(a,a')$ & $\widehat\se_\CR\{\widehat\tau_{j,\rmI}^\ancova(a,a')\}$ & Corollary \ref{coro:consistency-AN-I-ancova} & $\widehat\tau_{j,\rmT}(a,a')$ & $\widehat\se_\HC\{\widehat\tau_{j,\rmT}(a,a')\}$ & Theorem \ref{thm:consistency-AN-T} \\
					&&& $\widehat\tau_{j,\rmT}^\adj(a,a')$ & $\widehat\se_\HC\{\widehat\tau_{j,\rmT}^\adj(a,a')\}$ & Theorem \ref{thm:consistency-AN-T-cov} \\
					\hline
				\end{tabular}
			}
		\end{center}
	\end{table}
	\vspace{-0.2in}
	
	\subsection{Estimators using individual-level data and cluster-period averages}
	
	We first present additional results on the connections between estimators using individual-level data and cluster-period averages, summarizing them in Proposition \ref{prop:asy-vars-I-A}. Here, we explicitly write specific types of covariates in brackets after an estimator to emphasize the covariates used for baseline adjustment. All covariates are centered when included in the regression model. For example, $\widehat\tau_{j,\rmI}^\adj(a,a')[\bfX_{ijk}]$ denotes the regression estimator using individual-level data adjusted for $\bfX_{ijk}$ as covariates in the working model.
	\begin{proposition} \label{prop:asy-vars-I-A}
		For $j=1,\ldots,J$, assuming $a<a'\in\calA$, we have $\widehat\tau_{j,\rmI}(a,a')=\widehat\tau_{j,\rmA}(a,a')$, $\widehat\se_\CR^2\{\widehat\tau_{j,\rmI}(a,a')\}=\widehat\se_\HC^2\{\widehat\tau_{j,\rmA}(a,a')\}$, and $\widehat\tau_{j,\rmI}^\adj(a,a')[\bfC_{ij}]=\widehat\tau_{j,\rmA}^\adj(a,a')$, $\widehat\se_\CR^2\{\widehat\tau_{j,\rmI}^\adj(a,a')[\bfC_{ij}]\}=\widehat\se_\HC^2\{\widehat\tau_{j,\rmA}^\adj(a,a')\}$.
	\end{proposition}
	
	Without covariate adjustment, $\widehat\tau_{j,\rmI}(a,a')$ and $\widehat\tau_{j,\rmA}(a,a')$ are equivalent, and so are their corresponding variance estimators $\widehat\se_\CR^2\{\widehat\tau_{j,\rmI}(a,a')\}$ and $\widehat\se_\HC^2\{\widehat\tau_{j,\rmA}(a,a')\}$. This double equivalence extends to covariate-adjusted estimators if only cluster-level covariates $\bfC_{ij}$ are used for the estimator using individual-level data. This is consistent with the results given in \citet[Proposition 2]{Su2021} for parallel-arm CREs.
	
	\subsection{Estimators using individual-level data and scaled cluster-period totals}
	
	We compare $\widehat\tau_{j,\rmI}(a,a')$ and $\widehat\tau_{j,\rmT}(a,a')$, where by construction, $\widehat\tau_{j,\rmT}(a,a')$ is the Horvitz-Thompson estimator and is unbiased, while $\widehat\tau_{j,\rmI}(a,a')$ is the Haj\'ek estimator and is biased. Nevertheless, $\widehat\tau_{j,\rmI}(a,a')$ is consistent since its bias diminishes to zero as $I\rightarrow\infty$. Similar to \citet[\S 3.4]{Su2021} for the parallel-arm CREs, $\widehat\tau_{j,\rmI}(a,a')$ has a smaller asymptotic variance than $\widehat\tau_{j,\rmT}(a,a')$ because the residual $\widetilde Y_{ij\cdot}(a)-I\pi_{ij\cdot}\overline Y_{\cdot j\cdot}(a)$ in $\se^2\{\widehat\tau_{j,\rmI}(a,a')\}$ adjusts for normalized cluster-period weight $\pi_{ij\cdot}$, while the residual $\widetilde Y_{ij\cdot}(a)-\overline Y_{\cdot j\cdot}(a)$ in $\se^2\{\widehat\tau_{j,\rmI}(a,a')\}$ does not. More importantly, $\widehat\tau_{j,\rmT}(a,a')$ is not invariant to the location shift of the potential outcomes. That is, if we transform the potential outcomes to $m + Y_{ijk}(a)$ for $a\in\calA$ and some constant $m$, then $\widehat\tau_{j,\rmT}(a,a')$ will change but $\widehat\tau_{j,\rmI}(a,a')$ will not. Therefore, we do not recommend using $\widehat\tau_{j,\rmT}(a,a')$ without covariates in general. The shortcomings of $\widehat\tau_{j,\rmT}(a,a')$ can be fixed by leveraging covariate adjustment with cluster-level information $\bfC_{ij}$, for which a simple case is given by choosing a simple scalar covariate $C_{ij}=\pi_{ij\cdot}$. Proposition \ref{prop:asy-vars-I-T} summarizes the results comparing asymptotic variances of these estimators.
	
	\begin{proposition} \label{prop:asy-vars-I-T}
		For $j=1,\ldots,J$, assuming $a<a'\in\calA$, the asymptotic variances satisfy $\se^2\{\widehat\tau_{j,\rmT}^\adj(a,a')[\pi_{ij\cdot},\widetilde \bfX_{ij\cdot}]\}\leq \se^2\{\widehat\tau_{j,\rmT}^\adj(a,a')[\pi_{ij\cdot}]\}\leq \se^2\{\widehat\tau_{j,\rmI}(a,a')\}$, and $\se^2\{\widehat\tau_{j,\rmT}^\adj(a,a')[\pi_{ij\cdot},\allowbreak\widetilde \bfX_{ij\cdot}]\}\leq\se^2\{\widehat\tau_{j,\rmI}^\adj(a,a')\}.$ These inequalities hold asymptotically if the standard errors are replaced by the estimated ones.
	\end{proposition}
	
	The inequality $\se^2\{\widehat\tau_{j,\rmT}^\adj(a,a')[\pi_{ij\cdot},\widetilde \bfX_{ij\cdot}]\}\leq \se^2\{\widehat\tau_{j,\rmT}^\adj(a,a')[\pi_{ij\cdot}]\}$ is established following the techniques to prove the result that covariate adjustment will always improve the asymptotic variance under the independent data setting \citep{Lin2013}. The inequality $\se^2\{\widehat\tau_{j,\rmT}^\adj(a,a')[\pi_{ij\cdot}]\}\leq \se^2\{\widehat\tau_{j,\rmI}(a,a')\}$ can be obtained by first establishing the following expression $\sumi\widetilde\epsilon_{ij\cdot}(a)^2=\sumi\{\widetilde Y_{ij\cdot}(a)-I\pi_{ij\cdot}\overline Y_{\cdot j\cdot}(a)\}^2=\sumi\{\widetilde Y_{ij\cdot}(a)-\overline Y_{\cdot j\cdot}(a)-(\pi_{ij\cdot}-I^{-1})I\overline Y_{\cdot j\cdot}(a)\}^2$. Then, by the Frisch-Waugh-Lovell Theorem \citep{Frisch1933}, the quantity $\sumi r_{ij}(a)^2$ for $\se^2\{\widehat\tau_{j,\rmT}^\adj(a,a')[\pi_{ij\cdot}]\}$ in Theorem \ref{thm:consistency-AN-T-cov} is the residual sum of squares from the OLS fit of $\widetilde Y_{ij\cdot}(a)-\overline Y_{ij\cdot}(a)$ on $\pi_{ij\cdot}-I^{-1}$, which by definition is smaller than $\sumi\{\widetilde Y_{ij\cdot}(a)-\overline Y_{\cdot j\cdot}(a)-(\pi_{ij\cdot}-I^{-1})I\overline Y_{\cdot j\cdot}(a)\}^2$. Therefore, adjusting for $\pi_{ij\cdot}$ in the regression using scaled cluster-period totals improves the simple difference-in-means. Additionally, adjusting for scaled cluster-period totals of other covariates leads to greater asymptotic efficiency improvement. The same argument can be applied to obtaining the inequality $\se^2\{\widehat\tau_{j,\rmT}^\adj(a,a')[\pi_{ij\cdot},\widetilde \bfX_{ij\cdot}]\}\leq\se^2\{\widehat\tau_{j,\rmI}^\adj(a,a')\}$. Specifically, the quantity $\sumi r_{ij}(a)^2=\sumi\{\widetilde \epsilon_{ij\cdot}(a)-\widetilde\bfX_{ij\cdot}^c\bfgamma_{j,\rmI}(a)\}^2$ for $\se^2\{\widehat\tau_{j,\rmI}^\adj(a,a')\}$ in Theorem \ref{thm:consistency-AN-I-cov} can be expressed as $\sumi\{\widetilde Y_{ij\cdot}(a)-\overline Y_{\cdot j\cdot}(a)-(\pi_{ij\cdot}-I^{-1})I\overline Y_{\cdot j\cdot}(a)-\widetilde\bfX_{ij\cdot}^c\bfgamma_{j,\rmI}(a)\}^2$, which is greater than the residual sum of squares from the OLS fit of $\widetilde Y_{ij\cdot}(a)-\overline Y_{ij\cdot}(a)$ on $(\pi_{ij\cdot}-I^{-1},\widetilde\bfX_{ij\cdot}^c)$, that is, the quantity $\sumi r_{ij}(a)^2$ for $\se^2\{\widehat\tau_{j,\rmT}^\adj(a,a')[\pi_{ij\cdot},\widetilde\bfX_{ij\cdot}]\}$ in Theorem \ref{thm:consistency-AN-T-cov}. Therefore, adjusting for $(\pi_{ij\cdot},\widetilde \bfX_{ij\cdot})$ in the regression using scaled cluster-period totals improves the regression adjustment of $\bfX_{ijk}$ using the individual-level data, regarding asymptotic efficiency. 
	
	\subsection{Estimators using cluster-period averages and scaled cluster-period totals} \label{subsec:rec}
	
	We compare $\widehat\tau_{j,\rmA}(a,a')$ and $\widehat\tau_{j,\rmT}(a,a')$. Similar to $\widehat\tau_{j,\rmI}(a,a')$, $\widehat\tau_{j,\rmA}(a,a')$ is the Haj\'ek estimator and is consistent with its bias diminishing to zero as $I\rightarrow\infty$. The following Proposition \ref{prop:asy-vars-A-T} summarizes the results comparing asymptotic variances of estimators using cluster-period averages and scaled cluster-period totals.
	
	\begin{proposition} \label{prop:asy-vars-A-T}
		For $j=1,\ldots,J$, assuming $a<a'\in\calA$, the asymptotic variances satisfy
		$\se^2\{\widehat\tau_{j,\rmT}^\adj(a,a')[\pi_{ij\cdot},\pi_{ij\cdot}\bfC_{ij}]\}\leq \se^2\{\widehat\tau_{j,\rmT}^\adj(a,a')[\pi_{ij\cdot}]\}\leq \se^2\{\widehat\tau_{j,\rmA}(a,a')\}$, and $\se^2\{\widehat\tau_{j,\rmT}^\adj(a,a')[\pi_{ij\cdot},\allowbreak\pi_{ij\cdot}\bfC_{ij}]\}\leq\se^2\{\widehat\tau_{j,\rmA}^\adj(a,a')\}$. These inequalities hold asymptotically if the standard errors are replaced by the estimated ones.
	\end{proposition}
	
	Proposition \ref{prop:asy-vars-A-T} is obtained via similar arguments centered around the sum of squares and the Frisch-Waugh-Lovell Theorem used in deriving Proposition \ref{prop:asy-vars-I-T}. The inequality $\se^2\{\widehat\tau_{j,\rmT}^\adj(a,a')[\pi_{ij\cdot},\pi_{ij\cdot}\bfC_{ij}]\}\leq \se^2\{\widehat\tau_{j,\rmT}^\adj(a,a')[\pi_{ij\cdot}]\}$ follows the result that regression adjustment will always improve the asymptotic variance \citep{Lin2013}. The inequality $\se^2\{\widehat\tau_{j,\rmT}^\adj(a,a')[\pi_{ij\cdot}]\}\allowbreak\leq \se^2\{\widehat\tau_{j,\rmA}(a,a')\}$ is obtained from the fact that $\sumi\widetilde\epsilon_{ij\cdot}(a)^2=\sumi\{\widetilde Y_{ij\cdot}(a)-I\pi_{ij\cdot}\overline Y_{\cdot j\cdot}(a)\}^2=\sumi\{\widetilde Y_{ij\cdot}(a)-\overline Y_{\cdot j\cdot}(a)-(\pi_{ij\cdot}-I^{-1})I\overline Y_{\cdot j\cdot}(a)\}^2$, and $\sumi r_{ij}(a)^2$ for $\se^2\{\widehat\tau_{j,\rmT}^\adj(a,a')[\pi_{ij\cdot}]\}$ in Theorem \ref{thm:consistency-AN-T-cov} is the residual sum of squares from the OLS fit of $\widetilde Y_{ij\cdot}(a)-\overline Y_{ij\cdot}(a)$ on $\pi_{ij\cdot}-I^{-1}$, which by definition is smaller than $\sumi\{\widetilde Y_{ij\cdot}(a)-\overline Y_{\cdot j\cdot}(a)-(\pi_{ij\cdot}-I^{-1})I\overline Y_{\cdot j\cdot}(a)\}^2$. Therefore, adjusting for $\pi_{ij\cdot}$ improves the efficiency of simple difference-in-means, and adjusting for scaled cluster-period totals of additional covariates leads to a greater improvement in efficiency. The same argument can be applied to obtain the inequality $\se^2\{\widehat\tau_{j,\rmT}^\adj(a,a')[\pi_{ij\cdot},\pi_{ij\cdot}\bfC_{ij}]\}\leq\se^2\{\widehat\tau_{j,\rmA}^\adj(a,a')\}$, where the quantity $\sumi r_{ij}(a)^2=\sumi\{\widetilde \epsilon_{ij\cdot}(a)-I\pi_{ij\cdot}\bfC_{ij}^c\bfgamma_{j,\rmI}(a)\}^2$ for $\se^2\{\widehat\tau_{j,\rmA}^\adj(a,a')\}$ in Theorem \ref{thm:consistency-AN-A-cov} can be expressed as $\sumi\{\widetilde Y_{ij\cdot}(a)-\overline Y_{\cdot j\cdot}(a)-(\pi_{ij\cdot}-I^{-1})I\overline Y_{\cdot j\cdot}(a)-\pi_{ij\cdot}\bfC_{ij}^cI\bfgamma_{j,\rmA}(a)\}^2$, which is greater than the residual sum of squares from the OLS fit of $\widetilde Y_{ij\cdot}(a)-\overline Y_{ij\cdot}(a)$ on $(\pi_{ij\cdot}-I^{-1},\pi_{ij\cdot}\bfC_{ij}^c)$, i.e., the quantity $\sumi r_{ij}(a)^2$ for $\se^2\{\widehat\tau_{j,\rmT}^\adj(a,a')[\pi_{ij\cdot},\pi_{ij\cdot}\bfC_{ij}]\}$ in Theorem \ref{thm:consistency-AN-T-cov}. Therefore, adjusting for $(\pi_{ij\cdot},\pi_{ij\cdot} \bfC_{ij})$ improves the regression adjustment of $\bfC_{ij}$ using cluster-period averages, regarding asymptotic efficiency. 
	
	\subsection{Generalization to summary estimands}\label{sec:gen_compare}
	
	Results in Propositions \ref{prop:asy-vars-I-A} - \ref{prop:asy-vars-A-T} can be generalized to summary estimands. The generalization of Proposition \ref{prop:asy-vars-I-A} is apparent. For Propositions \ref{prop:asy-vars-I-T} and \ref{prop:asy-vars-A-T}, consider the example of comparing $\se^2(\widehat\theta_\rmT^\adj[\pi_{ij\cdot}])$ and $\se^2(\widehat\theta_\rmA^\adj)$, where $\se^2(\widehat\theta_\rmA^\adj)-\se^2(\widehat\theta_\rmT^\adj[\pi_{ij\cdot}])=\bfb_\theta^\top\{\Cov(\widehat\bftau_\rmA^\adj)-\Cov(\widehat\bftau_\rmT^\adj[\pi_{ij\cdot}])\}\bfb_\theta$. Using the arguments for Proposition \ref{prop:asy-vars-A-T}, the diagonal entries of $\Cov(\widehat\bftau_\rmA^\adj)$ are greater than the corresponding entries of $\Cov(\widehat\bftau_\rmT^\adj[\pi_{ij\cdot}])$. Since both $\Cov(\widehat\bftau_\rmA^\adj)$ and $\Cov(\widehat\bftau_\rmT^\adj[\pi_{ij\cdot}])$ are covariance matrices and, therefore, positive definite, the difference matrix $\Cov(\widehat\bftau_\rmA^\adj)-\Cov(\widehat\bftau_\rmT^\adj[\pi_{ij\cdot}])$ is positive semi-definite such that $\bfb_\theta^\top\{\Cov(\widehat\bftau_\rmA^\adj)-\Cov(\widehat\bftau_\rmT^\adj[\pi_{ij\cdot}])\}\bfb_\theta\geq 0$ for any $\bfb_\theta\in\bbR^{J^2(J+1)/2}$, suggesting that $\widehat\theta_\rmT^\adj[\pi_{ij\cdot}]$ is more efficient than $\widehat\theta_\rmA^\adj$ with $\se^2(\widehat\theta_\rmT^\adj[\pi_{ij\cdot}])\leq \se^2(\widehat\theta_\rmA^\adj)$. The following Corollary \ref{coro:asy-vars-sum-est} summarizes the results comparing asymptotic variances of estimators for summary estimands.
	
	\begin{corollary} \label{coro:asy-vars-sum-est}
		\begin{itemize}
			\item[(a)] For estimators using individual-level data and cluster-period averages, we have 
			$\widehat\theta_\rmI=\widehat\theta_\rmA$, $\widehat\se_\CR^2(\widehat\theta_\rmI)=\widehat\se_\HC^2(\widehat\theta_\rmA)$, $\widehat\theta_\rmI^\adj[\bfC_{ij}]=\widehat\theta_\rmA^\adj$, and $\widehat\se_\CR^2(\widehat\theta_\rmI^\adj[\bfC_{ij}])=\widehat\se_\HC^2(\widehat\theta_\rmA^\adj)$. 
			\item[(b)] For estimators using individual-level data and scaled cluster-period totals, we have 
			$\se^2(\widehat\theta_\rmT^\adj[\pi_{ij\cdot},\widetilde \bfX_{ij\cdot}])\leq \se^2(\widehat\theta_\rmT^\adj[\pi_{ij\cdot}])\leq \se^2(\widehat\theta_\rmI)$, and $\se^2(\widehat\theta_\rmT^\adj[\pi_{ij\cdot},\widetilde \bfX_{ij\cdot}])\leq\se^2(\widehat\theta_\rmI^\adj)$.
			\item[(c)] For estimators using cluster-period averages and scaled cluster-period totals, we have 
			$\se^2(\widehat\theta_\rmT^\adj[\pi_{ij\cdot},\pi_{ij\cdot}\bfC_{ij}])\leq \se^2(\widehat\theta_\rmT^\adj[\pi_{ij\cdot}])\leq \se^2(\widehat\theta_\rmA)$, and $\se^2(\widehat\theta_\rmT^\adj[\pi_{ij\cdot},\pi_{ij\cdot}\bfC_{ij}])\leq\se^2(\widehat\theta_\rmA^\adj)$.
		\end{itemize}
	\end{corollary}

	\section{Simulation studies}  \label{sec:sim}
	
	To illustrate the asymptotic properties, we conduct simulations to compare regression estimators and their standard errors using different data-generating processes. To focus ideas, in each simulation study, we estimate the DWATE estimands and the calendar time-weighted overall weighted treatment effects, $\OWTE^{cal}$.
	
	\subsection{Simulation study I} \label{subsec:sim-1}
	
\begin{figure}[htbp]
		\centering
		\begin{subfigure}{\textwidth}
			\centering
			\includegraphics[width=0.33\textwidth]{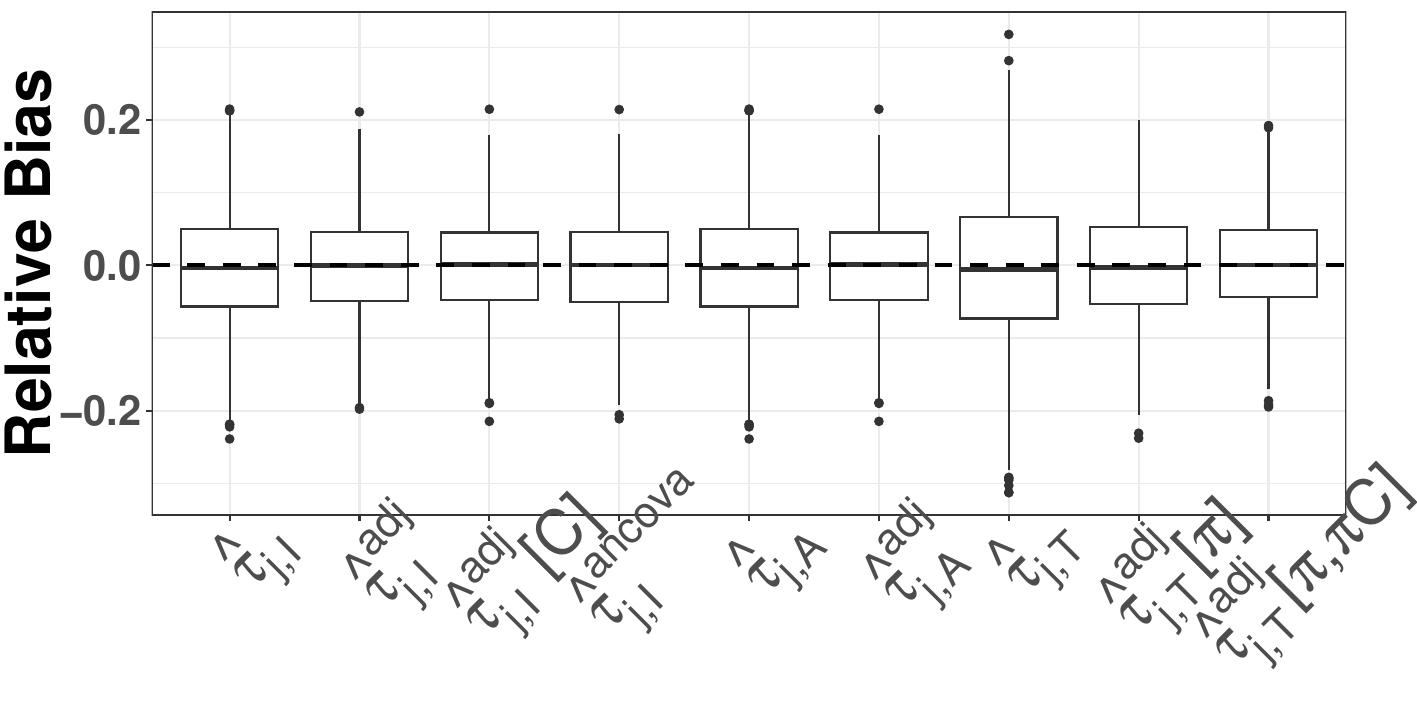}\hfill
			\includegraphics[width=0.33\textwidth]{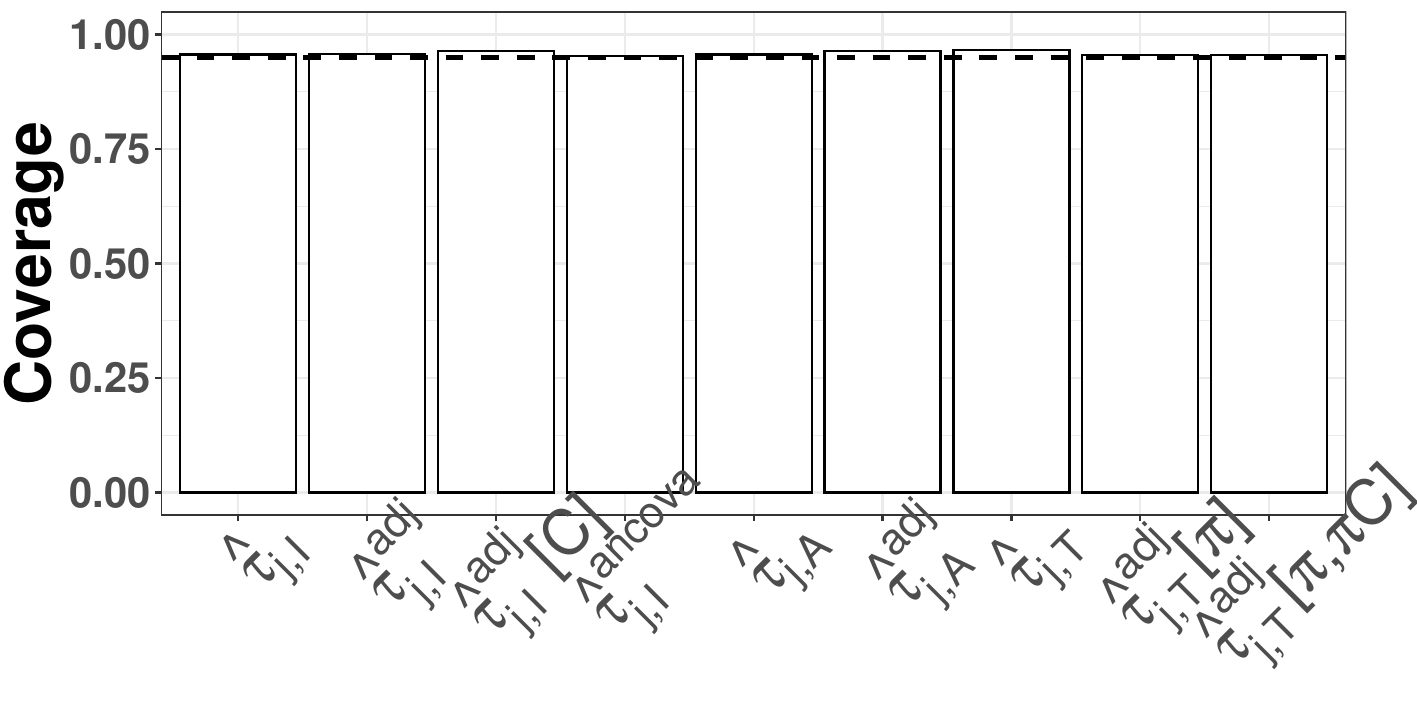}\hfill
			\includegraphics[width=0.33\textwidth]{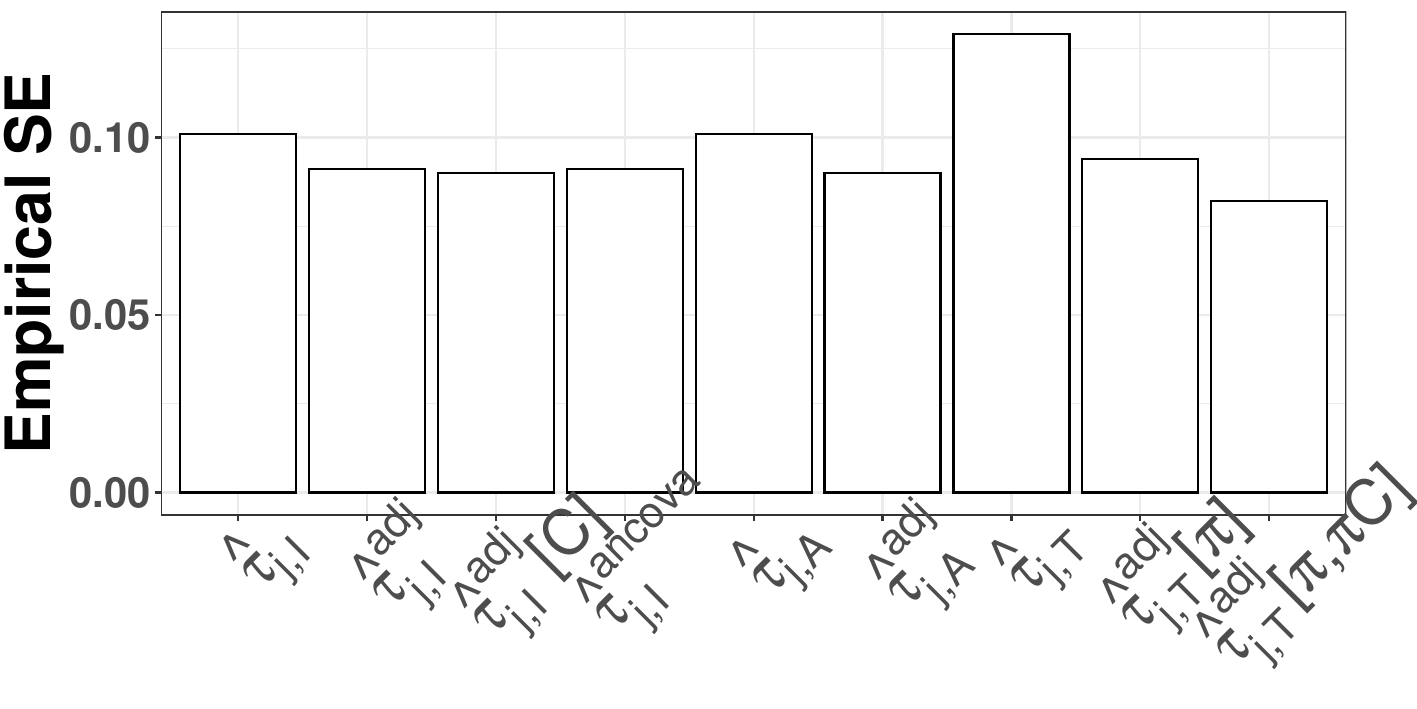}
			\caption{Simulation study I: comparing estimators and standard errors}
			\label{subfig:tau_11-sim-1}
		\end{subfigure}
		\\[4ex]
		\begin{subfigure}{\textwidth}
			\centering
			\includegraphics[width=0.33\textwidth]{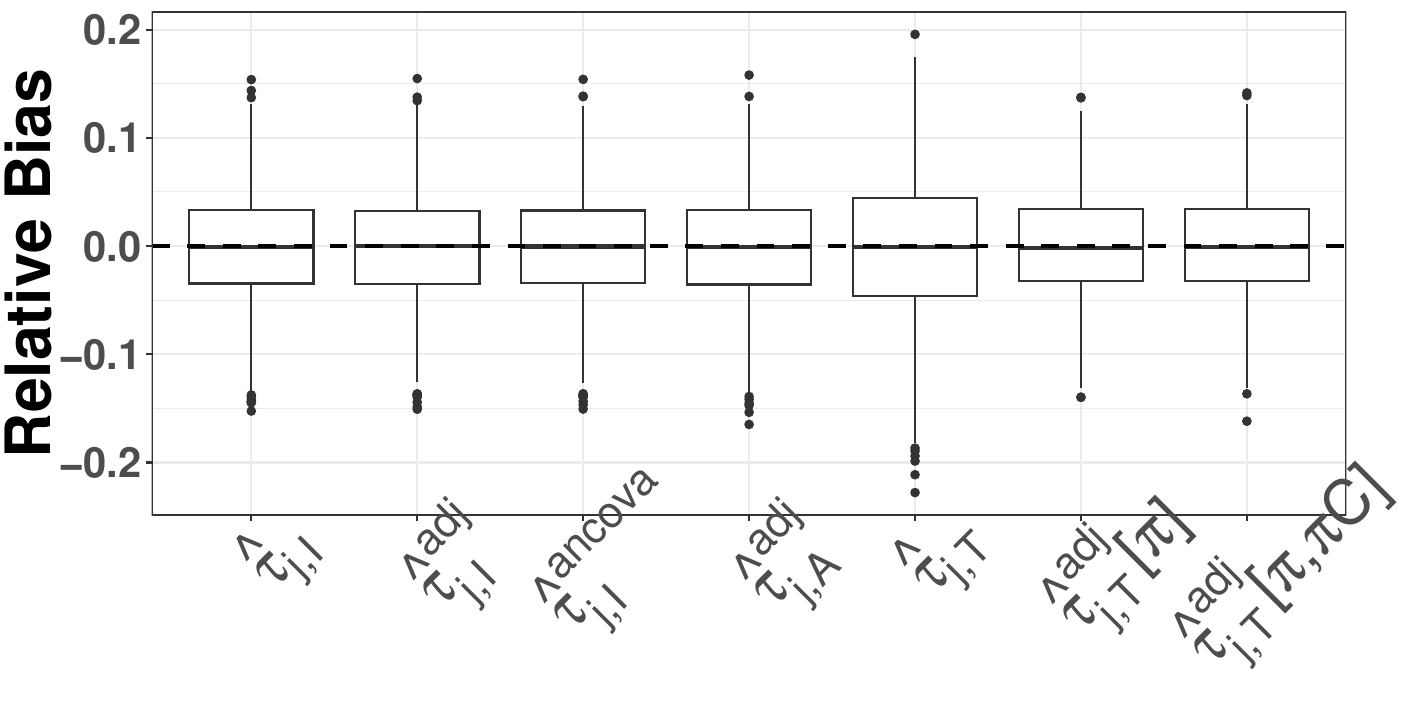}\hfill
			\includegraphics[width=0.33\textwidth]{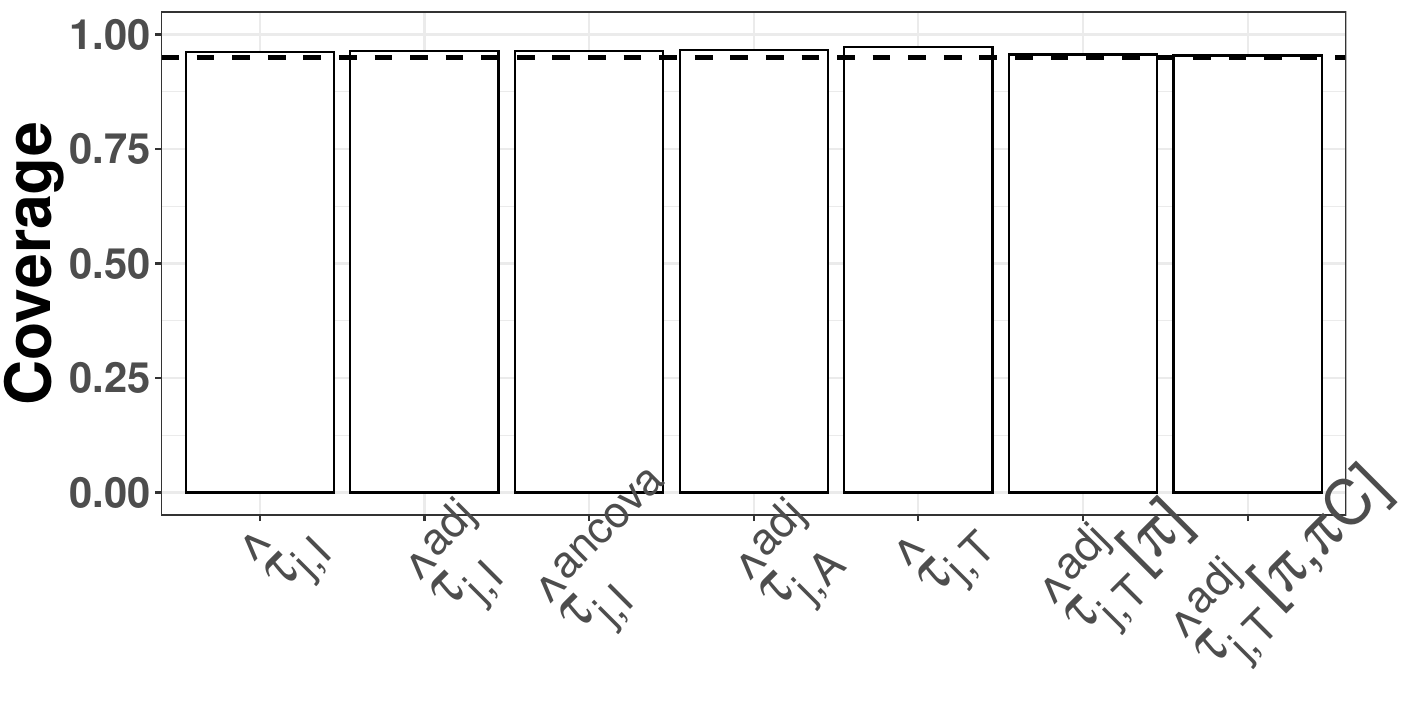}\hfill
			\includegraphics[width=0.33\textwidth]{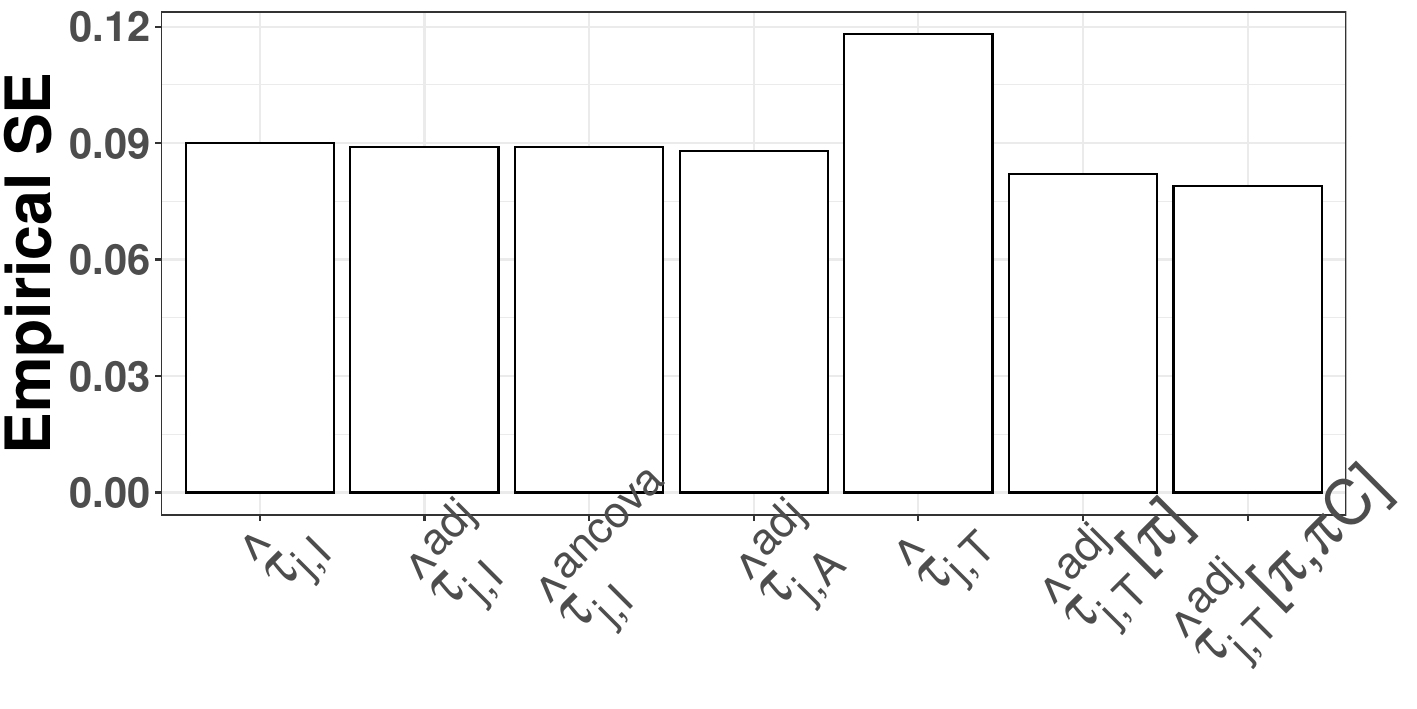}
			\caption{Simulation study II: uninformative covariates}
			\label{subfig:tau_11-sim-2}
		\end{subfigure}
		\caption{Relative bias, coverage percentages of 95\% CIs, and empirical standard errors for $\tau_1(1,\infty)$ from simulation studies I and II.}
		\label{fig:tau_11}
	\end{figure}

\begin{figure}[htbp]
		\centering
		\begin{subfigure}{\textwidth}
			\centering
			\includegraphics[width=0.33\textwidth]{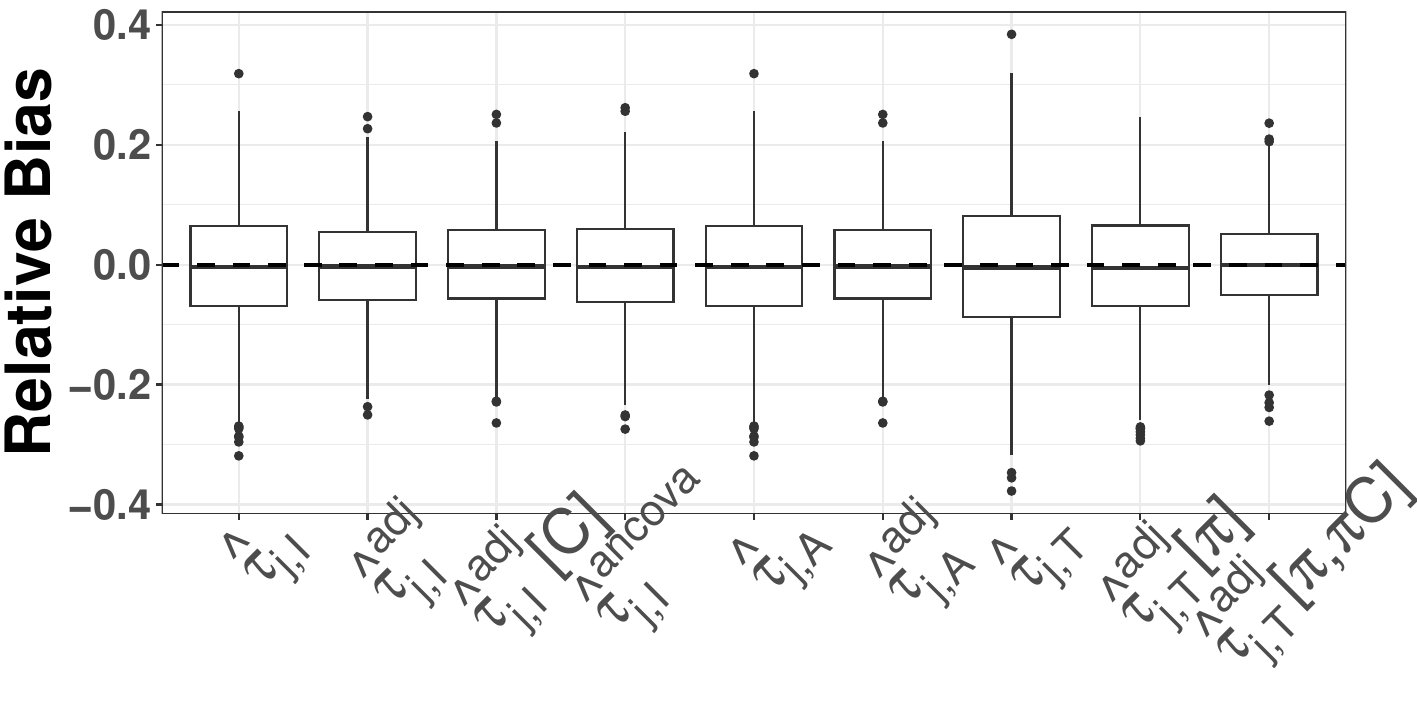}\hfill
			\includegraphics[width=0.33\textwidth]{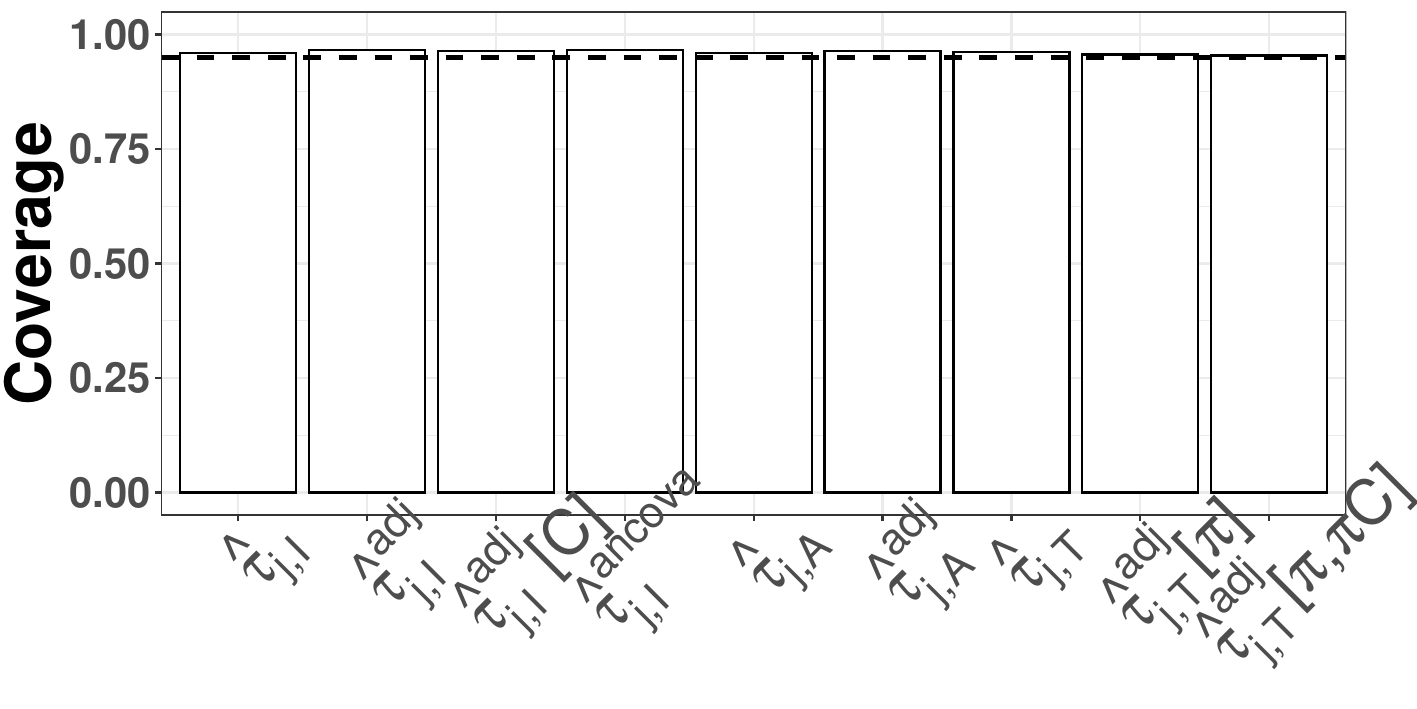}\hfill
			\includegraphics[width=0.33\textwidth]{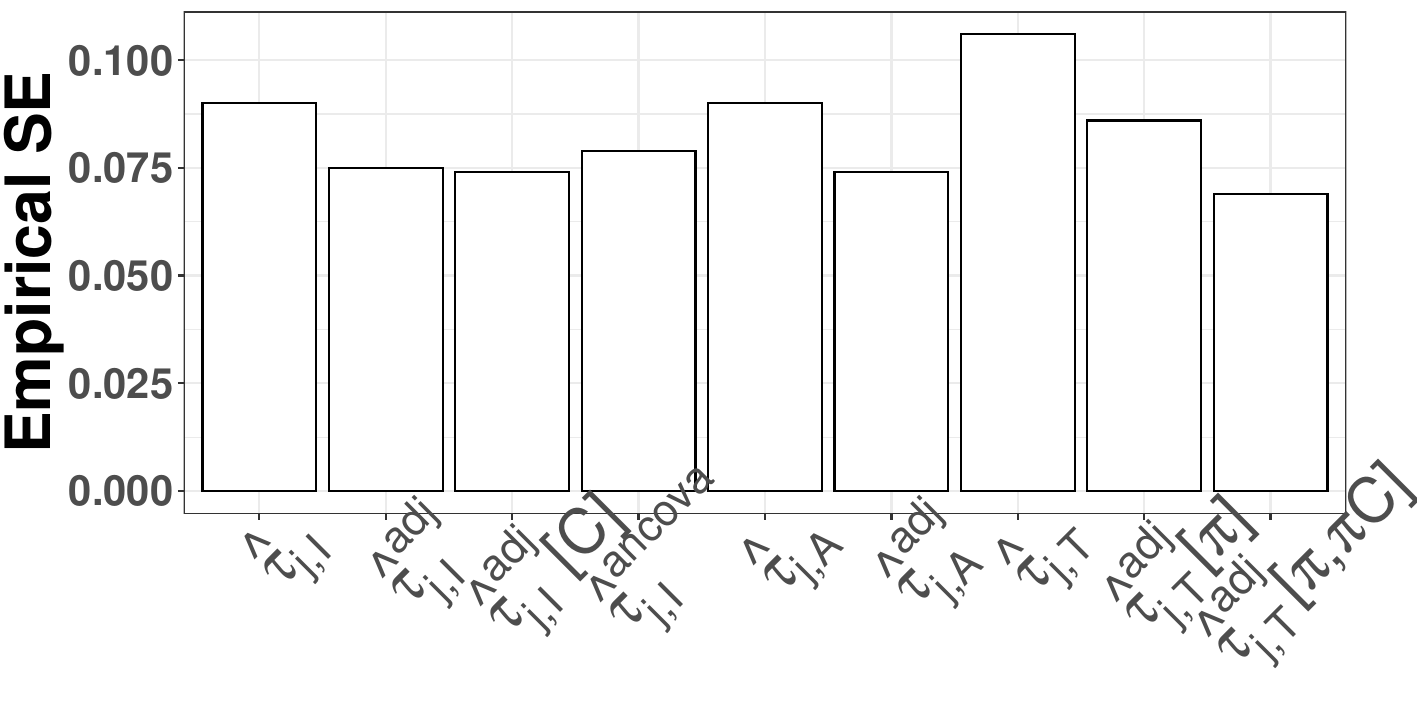}
			\caption{Simulation study I: comparing estimators and standard errors}
			\label{subfig:sum_est-sim-1}
		\end{subfigure}
		\\[4ex]
		\begin{subfigure}{\textwidth}
			\centering
			\includegraphics[width=0.33\textwidth]{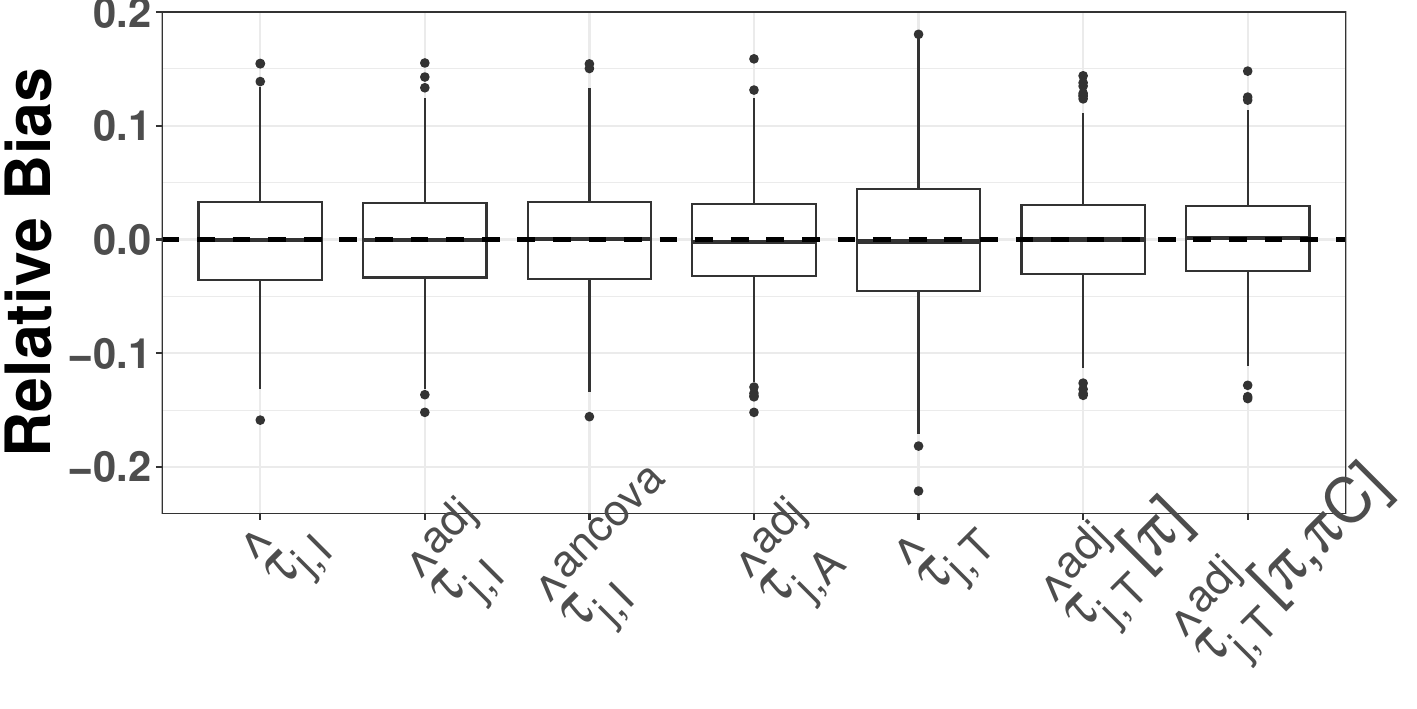}\hfill
			\includegraphics[width=0.33\textwidth]{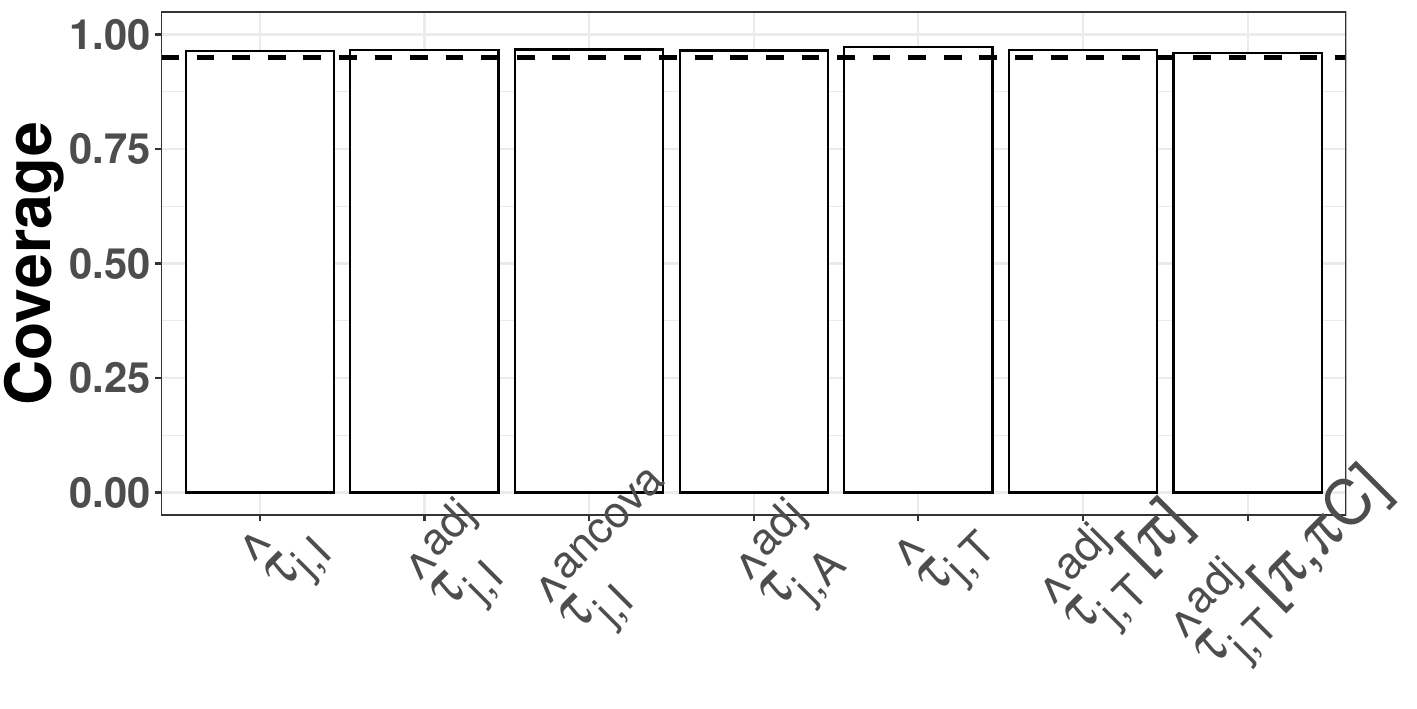}\hfill
			\includegraphics[width=0.33\textwidth]{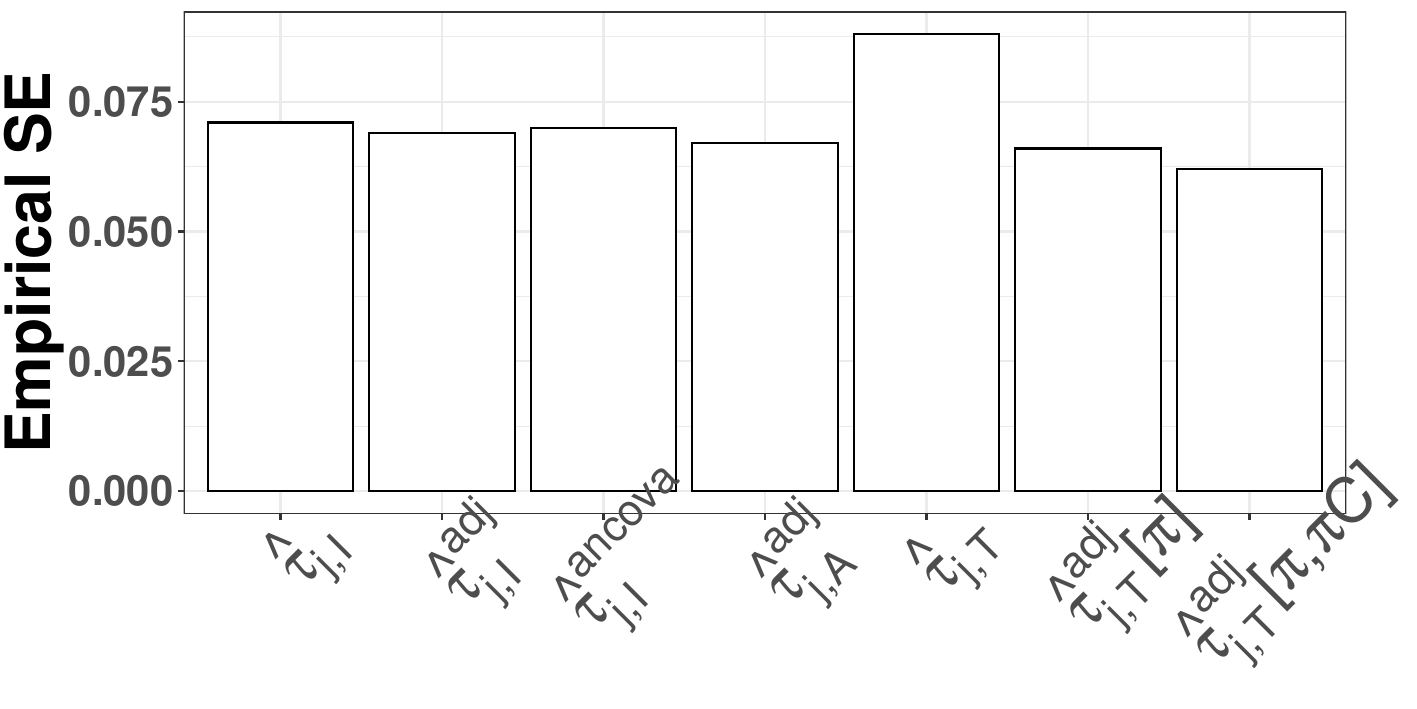}
			\caption{Simulation study II: uninformative covariates}
			\label{subfig:sum_est-sim-2}
		\end{subfigure}
		\caption{Relative bias, coverage percentages of 95\% CIs, and empirical standard errors for $\OWTE^{cal}$ from simulation studies I and II.}
		\label{fig:sum_est}
	\end{figure}

	Consider an SR-CRE with $I=260$ clusters and $J=2$ periods, and the treatment adoption times $\calA=\{1,2,\infty\}$ with $q(a)=1/3$ for $a\in\calA$. Let the individual weight $w_{ijk}=1$, and we generate $N_{ij}$ from the uniform distribution $\calU(5200/(jI) \times 0.6, 5200/(jI) \times 1.4)$. The potential outcomes are generated from the following model: $(Y_{ijk}(a)|X_{ijk})\sim\calN(f_a(i,j,X_{ijk}),1)$, with $f_1(i,j,X_{ijk})=2N_{ij}I/N_j+(X_{ijk}^c)^3+\zeta_{ij}$, $f_2(i,j,X_{ijk}) =\sqrt{N_{ij}}I/N_j (X_{ijk}^c)^4+\log|X_{ijk}^c|+\zeta_{ij}$, and $f_\infty(i,j,X_{ijk}) =i/I+(X_{ijk}^c)^2+\zeta_{ij}$, where $X_{ijk}\sim ij/I + \calU(-1,1)$ and the cluster-period random effect $\zeta_{ij}\sim\calN(0,0.2)$. This SR-CRE setting leads to six DWATE estimands, i.e., $\tau_1(1,\infty)$, $\tau_1(2,\infty)$, $\tau_1(1,2)$, $\tau_2(1,\infty)$, $\tau_2(2,\infty)$, and $\tau_2(1,2)$, where the estimation results for $\tau_1(1,\infty)$ are presented in Figure \ref{fig:tau_11}(a), and the rest are deferred to Section S7 of the Online Supplement. The estimation results for $\OWTE^{cal}$ are given in Figure \ref{fig:sum_est}(a). The simulation results are obtained via 1,000 replications.
	
	We compare nine estimators, using individual-level data: $\widehat\tau_{j,\rmI}$, $\widehat\tau_{j,\rmI}^\adj$, $\widehat\tau_{j,\rmI}^\adj[C]$, and $\widehat\tau_{j,\rmI}^\ancova$; using cluster-period averages: $\widehat\tau_{j,\rmA}$ and $\widehat\tau_{j,\rmA}^\adj$; and using scaled cluster-period totals: $\widehat\tau_{j,\rmT}$, $\widehat\tau_{j,\rmT}^\adj[\pi]$, and $\widehat\tau_{j,\rmT}^\adj[\pi,\pi C]$. For estimators with covariate adjustment, $\widehat\tau_{j,\rmI}^\adj$ and $\widehat\tau_{j,\rmI}^\ancova$ adjust for the centered covariate $X_{ijk}^c$, $\widehat\tau_{j,\rmI}^\adj[C]$ and $\widehat\tau_{j,\rmA}^\adj$ adjust for the cluster-level centered covariate $\overline X_{ij}^c$, $\widehat\tau_{j,\rmT}^\adj[\pi]$ adjusts for the centered version of the normalized cluster-period weight $\pi_{ij\cdot}=N_{ij}/N_j$, and $\widehat\tau_{j,\rmT}^\adj[\pi,\pi C]$ adjusts for the centered version of $(\pi_{ij\cdot},\pi_{ij\cdot}C_{ij})^\top$, where $C_{ij}=I\overline X_{ij}$. Under this setting, $\widehat\tau_{j,\rmT}^\adj[\pi,\pi C]$ is equivalent to $\widehat\tau_{j,\rmT}^\adj[\pi,\widetilde X]$ which adjusts for the centered $(\pi_{ij\cdot},\widetilde X_{ij})^\top$ where $\widetilde X_{ij} = I\pi_{ij\cdot}\overline X_{ij}=\pi_{ij\cdot}C_{ij}$. 
	
	The left panel of Figure \ref{fig:tau_11}(a) presents the relative bias. The nine estimators all exhibit little to no bias, and more specifically, $\widehat\tau_{j,\rmI}$ and $\widehat\tau_{j,\rmI}^\adj[C]$ are numerically equivalent to $\widehat\tau_{j,\rmA}$ and $\widehat\tau_{j,\rmA}^\adj$, respectively, confirming Proposition \ref{prop:asy-vars-I-A}. The middle panel of Figure \ref{fig:tau_11}(a) shows the coverage percentages of the 95\% confidence intervals (CIs), and the results confirm the conservativeness of the CR and HC variance estimators. The right panel of Figure \ref{fig:tau_11}(a) gives the empirical standard errors of these estimators. The empirical standard errors of $\widehat\tau_{j,\rmI}$ and $\widehat\tau_{j,\rmI}^\adj[C]$ are numerically equivalent to those of $\widehat\tau_{j,\rmA}$ and $\widehat\tau_{j,\rmA}^\adj$, respectively. The average estimated standard errors of $\widehat\tau_{j,\rmI}$ and $\widehat\tau_{j,\rmI}^\adj[C]$ are also numerically equivalent to those of $\widehat\tau_{j,\rmA}$ and $\widehat\tau_{j,\rmA}^\adj$, respectively (given in Web Table 2 in Section S8 of the Online Supplement), further supporting Proposition \ref{prop:asy-vars-I-A}. The empirical standard error of $\widehat\tau_{j,\rmT}^\adj[\pi,\pi C]$ is smaller than that of $\widehat\tau_{j,\rmT}^\adj[\pi]$, confirming the first inequality in Propositions \ref{prop:asy-vars-I-T} and \ref{prop:asy-vars-A-T}. The empirical standard error of $\widehat\tau_{j,\rmT}^\adj[\pi]$ is smaller than that of $\widehat\tau_{j,\rmI}$, confirming the second inequality in Propositions \ref{prop:asy-vars-I-T} and \ref{prop:asy-vars-A-T}. Last but not least, the empirical standard error of $\widehat\tau_{j,\rmT}^\adj[\pi,\pi C]$ is smaller than those of $\widehat\tau_{j,\rmI}^\adj$ and $\widehat\tau_{j,\rmA}^\adj$, confirming the third inequality in Propositions \ref{prop:asy-vars-I-T} and \ref{prop:asy-vars-A-T}. The same patterns can be found in Figure \ref{fig:sum_est} when estimating the summary estimand $\OWTE^{cal}$ using regression estimators.

	\subsection{Simulation study II} \label{subsec:sim-2}
	
	Section \ref{subsec:sim-1} shows that adjusting for covariates could improve the performance of estimators when the covariates are informative of the potential outcomes. In the second simulation study, we further investigate this effect from an alternative perspective where the potential outcomes are generated from a process with uninformative covariates. Specifically, we modify the data-generating process for the potential outcomes to: $Y_{ijk}(a)\sim\calN(f_a(i,j),1)$, with $f_1(i,j)=2N_{ij}I/N_j+\zeta_{ij}$, $f_2(i,j) =\sqrt{N_{ij}}I/N_j+\zeta_{ij}$, and $f_\infty(i,j) =i/I+\zeta_{ij}$, where $\zeta_{ij}\sim\calN(0,0.2)$. The estimation results for $\tau_1(1,\infty)$ are presented in Figure \ref{fig:tau_11}(b), and the rest are available in Web Table 3 in Section S8 of the Online Supplement. The results for $\OWTE^{cal}$ are given in Figure \ref{fig:sum_est}(b). We compare seven estimators, with $\widehat\tau_{j,\rmI}^\adj[C]$ and $\widehat\tau_{j,\rmA}$ removed because of their respective equivalency to $\widehat\tau_{j,\rmA}^\adj$ and $\widehat\tau_{j,\rmI}$.
	
	Comparing the relative bias, coverage percentages, and empirical standard errors of $\widehat\tau_{j,\rmI}$, $\widehat\tau_{j,\rmI}^\adj$, and $\widehat\tau_{j,\rmI}^\ancova$, we observe that, though adjusting for covariates does not lower the efficiency of the estimator, its improvement is marginal when the covariates are uninformative of the potential outcomes. For estimators using scaled cluster-period totals, $\widehat\tau_{j,\rmT}^\adj[\pi]$ and $\widehat\tau_{j,\rmT}^\adj[\pi,\pi C]$ perform better than $\widehat\tau_{j,\rmT}$ because the normalized cluster-period weight $\pi_{ij\cdot}$ is informative of the potential outcomes. The same patterns hold in Figure \ref{fig:sum_est}(b) for $\OWTE^{cal}$.
	
	\subsection{Additional simulation studies}
	
	Additional simulations are to further illustrate properties of the proposed estimators. In simulation study III, we provide a counterexample showing that covariate adjustment does not necessarily improve the performance of $\widehat\tau_{j,\rmI}$ and $\widehat\tau_{j,\rmA}$. In simulation study IV, we investigate the finite-sample performance of the estimators with a large but not dominant cluster. Results from simulation studies III and IV are presented in Sections S7 and S8 of the Online Supplement. Complementary simulations are carried out under the same data-generating processes, but with $I=60$ to examine finite-sample performance. The results of these simulations are presented in Section S8 of the Online Supplement and are similar.

	\section{Data example} \label{sec:data}
	
	We illustrate the proposed methods by analyzing the Acute Coronary Syndrome Quality Improvement in Kerala (ACS QUIK) trial data \citep{Huffman2018}. The ACS QUIK trial is a cross-sectional SR-CRE conducted to evaluate the effect of a quality improvement toolkit (the intervention) on health outcomes in patients with acute myocardial infarction. The cluster-level treatment was rolled out to participating hospitals; in the first four-month period, no hospitals received the toolkit; in the last four-month period, all hospitals received the toolkit, and the entire rollout consisted of the four four-month periods in between.
	
	We focus our analysis on the four rollout periods. Following the description of the design given in Section \ref{sec:design}, we have $\calA=\{1,2,3,4,\infty\}$, where we regard clusters that were given the tool kit in the last four-month period as never treated during the rollout, with 12, 12, 11, 11, and 11 clusters randomized to these treatment adoption times respectively. The numbers of individuals involved in each of the four rollout periods are 3,309, 3,515, 3,852, and 4,040, with average cluster-period sizes of 58.05, 61.67, 67.58, and 70.88, and standard deviations of cluster-period size distributions are 69.76, 69.56, 68.26, and 67.31. We study the binary outcome of the 30-day major adverse cardiovascular events, including death, reinfarction, stroke, and major bleeding. In the analysis, we consider three covariates that are the main risk factors: age, gender (equals one for males and zero for females), 
	and systolic blood pressure (SBP) for possible covariate adjustments. For illustration, we address two estimands defined on the causal risk difference scale---the overall weighted treatment effect (simple average), $\OWTE^{sim}$, and the overall anticipated weighted treatment effect (simple average), $\OAWTE^{sim}$, both defined at the end of Section \ref{sec:design}. 
	
	\begin{table}[htbp] %***
		\caption{Results from analyzing the ACS QUIK trial data.}\label{tab:data-1}
		% \vspace{-0.5cm}
		\begin{center}
			% \resizebox{\linewidth}{!}{
				\begin{tabular}{cc rrrrrrc} 
					\hline \\[-2ex]
					& & \multicolumn{1}{c}{$\widehat\tau_{j,\rmI}$} & \multicolumn{1}{c}{$\widehat\tau_{j,\rmI}^\adj$} & \multicolumn{1}{c}{$\widehat\tau_{j,\rmI}^\ancova$} & \multicolumn{1}{c}{$\widehat\tau_{j,\rmA}^\adj$} & \multicolumn{1}{c}{$\widehat\tau_{j,\rmT}$} & \multicolumn{1}{c}{$\widehat\tau_{j,\rmT}^\adj[\pi]$} & \multicolumn{1}{c}{$\widehat\tau_{j,\rmT}^\adj[\pi,\pi C]$} \\[0.8ex]
					\hline\\[-2ex]
					\multicolumn{9}{c}{$w_{ijk}=1$}\\[0.8ex]
					\hline\\[-2ex]
					\multirow{2}{*}{$\OWTE^{sim}$} & est & .0050 & .0050 & .0050 & \textbf{.0087} & .0050 & .0059 & .0075 \\
					& se & .0030 & .0030 & .0030 & \textbf{.0025} & .0028 & .0039 & .0042 \\
					\hline\\[-2ex]
					\multirow{2}{*}{$\OAWTE^{sim}$} & est & \textbf{.0256} & \textbf{.0256} & \textbf{.0256} & \textbf{.0363} & .0290 & \textbf{.0707} & \textbf{.0959} \\
					& se & \textbf{.0119} & \textbf{.0117} & \textbf{.0121} & \textbf{.0088} & .0159 & \textbf{.0214} & \textbf{.0152} \\
					\hline\\[-2ex]
					\multicolumn{9}{c}{$w_{ijk}=N_{ij}^{-1}$}\\[0.8ex]
					\hline\\[-2ex]
					\multirow{2}{*}{$\OWTE^{sim}$} & est & $-$.0197 & $-$.0127 & $-$.0169 & .0210 & $-$.0198 & $-$.0185 & .0236 \\
					& se & .0228 & .0147 & .0209 & .0117 & .0246 & .0247 & .0127 \\
					\hline\\[-2ex]
					\multirow{2}{*}{$\OAWTE^{sim}$} & est & $-$.0253 & $-$.0135 & $-$.0227 & .0343 & $-$.0267 & $-$.0276 & .0369 \\
					& se & .0341 & .0231 & .0318 & .0181 & .0367 & .0368 & .0195 \\
					\hline\\[-2ex]
					\multicolumn{9}{l}{\footnotesize Bold font indicates the 95\% confidence interval does not cover zero.}
				\end{tabular}
				% }
		\end{center}
	\end{table}
	
	We consider two sets of individual weights, i.e., $w_{ijk}=1$ and $w_{ijk}=N_{ij}^{-1}$. The former leads to the DWATE estimand in period $j$ defined as the pooled average, $\tau_j(a,a')=N_j^{-1}\sumi\sumk\{Y_{ijk}(a)-Y_{ijk}(a')\}$, and the latter accounts for the imbalance in cluster sizes and defined an average of cluster means, $\tau_j(a,a')=I^{-1}\sumi\{\overline Y_{ij\cdot}(a)-\overline Y_{ij\cdot}(a')\}$ with $\overline Y_{ij\cdot}(a)=N_{ij}^{-1}\sumi Y_{ijk}(a)$ and $\overline Y_{ij\cdot}(a')=N_{ij}^{-1}\sumi Y_{ijk}(a')$. We estimate these estimands using the seven methods developed previously, using individual data: $\widehat\tau_{j,\rmI}$, $\widehat\tau_{j,\rmI}^\adj$, and $\widehat\tau_{j,\rmI}^\ancova$; using cluster-period averages: $\widehat\tau_{j,\rmA}^\adj$; and using scaled cluster-period totals: $\widehat\tau_{j,\rmT}^\adj$, $\widehat\tau_{j,\rmT}^\adj[\pi]$, and $\widehat\tau_{j,\rmT}^\adj[\pi,\pi C]$, with results presented in Table \ref{tab:data-1}. Under the weight $w_{ijk}=1$, only $\widehat\tau_{j,\rmA}^\adj$ returned with an estimate of the $\OWTE^{sim}$ that is significantly greater than zero, while six out of seven estimators yielded statistically significant estimates of the $\OAWTE^{sim}$. Under the weight $w_{ijk}=N_{ij}^{-1}$, no estimators yielded statistically significant estimates of $\OWTE^{sim}$ or $\OAWTE^{sim}$. However, under the weight $w_{ijk}=1$, because of the large variation in cluster-period sizes (coefficients of variation of cluster-period sizes are near one for all four periods), the regularity condition (C2) in Section S2 of the Online Supplement, requiring no clusters to have weights significantly dominant over others in each period, is likely violated. This could result in the asymptotic normality of the regression estimators and the conservativeness of variance estimators both being compromised, as demonstrated by the simulation results in Section S8 of the Online Supplement when a large cluster (cluster weight imbalance) exists and the simulation results in \citet[\S6.5]{Su2021} when a dominant cluster is present in a parallel-arm CRE. This situation can be avoided under the weight $w_{ijk}=N_{ij}^{-1}$ because it yields identical cluster weights, satisfying the aforementioned regularity condition (C2) and resulting in more trustworthy variance estimates. Therefore, combining these results, it could be concluded that the toolkit intervention did not have a significant treatment or placebo effect in reducing the risk of 30-day major adverse cardiovascular events during the rollout.

	\section{Concluding remarks} \label{sec:conc}
	
	We provide a comprehensive treatment of causal inference for the class of staggered rollout cluster randomized experiments, including a class of causal estimands accounting for adoption and calendar times, as well as regression estimators using different levels of data. To pursue our estimands, in each calendar period, we can first compare the difference between any pair of treatment adoption times by estimating the DWATE estimand. Other collections of estimands, including weighted average treatment effects and anticipated weighted average treatment effects (and examples listed in Section S1 of the Online Supplement), are linear combinations of the DWATEs. Regression estimators using individual-level data, cluster-period averages, and scaled cluster-period totals are formally characterized, which, under the design-based perspective, are model-assisted. We established the consistency and asymptotic normality for each regression estimator and proved that the CR and HC variance estimators are asymptotically conservative. Relatedly, our results on the conservativeness of the HC variance estimator also provide a rigorous theoretical foundation for its adoption in design-based analyses of SR-IREs \citep{Roth2023}. Through analytical comparisons, we found that regression estimators using individual-level data and cluster-period averages can be suboptimal, whereas the estimator using scaled cluster-period totals and adjusting for covariates is more efficient, with a smaller asymptotic variance. This implication, in fact, also offers novel insights into the analysis of cluster randomized DID. That is, if a cluster randomized DID design only has access to cluster-period level data, one should opt for collecting cluster-period totals rather than cluster-period averages, thereby making the analysis asymptotically more efficient. Our findings generalize those previously developed for much simpler parallel-arm CREs in \citet{Su2021} to a more complex class of SR-CREs. This generalization, however, is not trivial because our asymptotic results are established under the randomization distribution induced by assigning clusters to different adoption times and jointly for a finite-dimensional vector of DWATE estimands. The conservativeness of the CR and HC variance estimators, hence, is in the sense of L\"{o}wner ordering and holds for inference about any linear combinations of the DWATE estimands.
	
	We acknowledge that our asymptotic theory assumes the number of clusters approaches infinity, with bounded cluster-period sizes and a fixed number of calendar periods. This asymptotic regime is standard and follows the ones used in \citet{Middleton2015,Su2021,Chen2023}. Nevertheless, with SR-CREs, this is not the only possible asymptotic regime, and it will be interesting to develop complementary asymptotic theories under the alternative regime where the number of clusters stays fixed. Still, the cluster size approaches infinity or both approach infinity, reminiscent of those studied in \citet{xie2003asymptotics} for generalized estimating equation estimators.

	\section*{Acknowledgement}
	%\vspace{-0.3cm}
	Research in this article was supported by a Patient-Centered Outcomes Research Institute Award\textsuperscript{\textregistered} (PCORI\textsuperscript{\textregistered} Award ME-2022C2-27676). The statements presented are solely the responsibility of the authors and do not necessarily represent the official views of PCORI\textsuperscript{\textregistered}, its Board of Governors, or the Methodology Committee. We thank Hao Wang for helpful discussions and assistance with the data example.

	%\newpage
	\singlespacing
	\bibliographystyle{jasa3}
	\bibliography{MA_SR-CRE}

\end{document}